\documentclass[12pt]{article}

\def\ltwid{\mathrel{\raise.3ex\hbox{$<$\kern-.75em\lower1ex\hbox{$\sim$}}}}
\def\gtwid{\mathrel{\raise.3ex\hbox{$>$\kern-.75em\lower1ex\hbox{$\sim$}}}}
\def\comp{{\rm C}\llap{\vrule height7.1pt width1pt depth-.4pt\phantom t}}

\def\square{\kern1pt\vbox{\hrule height 1.2pt\hbox{\vrule width 1.2pt\hskip 3pt
   \vbox{\vskip 6pt}\hskip 3pt\vrule width 0.6pt}\hrule height 0.6pt}\kern1pt}

\begin{document}

\begin{titlepage}

\begin{flushright}
ITP-UU-12/09 \\ SPIN-12/08 \\ UFIFT-QG-11-04
\end{flushright}

\begin{center}
{\bf Issues Concerning Loop Corrections to the Primordial Power Spectra}
\end{center}

\begin{center}
S. P. Miao$^*$
\end{center}

\begin{center}
\it{Institute for Theoretical Physics \& Spinoza Institute, Utrecht
University \\ Leuvenlaan 4, Postbus 80.195, 3508 TD Utrecht, NETHERLANDS}\\
\end{center}

\begin{center}
R. P. Woodard$^{\dagger}$
\end{center}

\begin{center}
\it{Department of Physics, University of Florida \\
Gainesville, FL 32611, UNITED STATES}
\end{center}

\begin{center}
ABSTRACT
\end{center}
We expound ten principles in an attempt to clarify the debate over
infrared loop corrections to the primordial scalar and tensor power
spectra from inflation. Among other things we note that existing
proposals for nonlinear extensions of the scalar fluctuation field
$\zeta$ introduce new ultraviolet divergences which no one understands
how to renormalize. Loop corrections and higher correlators of these
putative observables would also be enhanced by inverse powers of the
slow roll parameter $\epsilon$. We propose an extension which should
be better behaved.

\begin{flushleft}
\end{flushleft}

\begin{flushleft}
$^*$ e-mail: S.Miao@uu.nl \\
$^{\dagger}$ e-mail: woodard@phys.ufl.edu
\end{flushleft}

\end{titlepage}

\section{Introduction}

The power spectra of primordial tensor \cite{Starobinsky} and scalar
\cite{Mukhanov} perturbations from inflation have assumed a crucial
place in fundamental theory because they describe the first
observable quantum gravitational effects. At present only the scalar
power spectrum has been observed, and without the sensitivity to
resolve even the first loop correction \cite{WMAP}. The correction
from each additional loop is suppressed by a factor of Newton's
constant $G$ times the square of the inflationary Hubble parameter
$H$. Assuming single scalar inflation, and using the measured value
of the scalar power spectrum and the upper bound on the
tensor-to-scalar ratio, one can conclude that $G H^2 \ltwid
10^{-10}$ \cite{KOW}. This would seem to be a crushing suppression
factor but it has been pointed out that the sensitivity to resolve
one loop corrections might be achieved by measuring the matter power
spectrum to very large redshifts \cite{21cm}. Realizing this
possibility would require a unique model of inflation, and an
enormous amount of work to untangle the primordial signal from late
time effects, but the first steps have already been taken
\cite{first21}.

In a situation like this we are obviously searching for every
conceivable source of enhancement in the theoretical signal. So it
is only natural that cosmologists and fundamental theorists have
been drawn to consider quantum infrared effects which formally, and
for the most naive extensions of the tree order observables, give
rise to infrared {\it divergent} loop corrections \cite{IRstudies}.
Of course no one believes the result is infinite, but the hope has
sometimes been expressed that the small loop counting parameter of
$G H^2$ might be partially compensated by a large infrared cutoff.
On the other hand, infrared effects derive from fields which are
nearly constant, and {\it exactly} constant graviton fields are pure
gauge. This has led to a countervailing argument that the apparent
infrared sensitivity of straightforward definitions of the ``power
spectrum'' can and should be eliminated by employing a gauge
invariant operator to represent the strength of primordial
fluctuations \cite{gauge}.

We incline to the view that the infrared divergence is pure gauge
for gravitons, but it disturbs us that this is being confused with
finite infrared effects which should be physical. Several other
points in the debate also seem to be unfortunate. In an effort to
clarify the situation we have identified ten principles which are
presented in section 3, after a review of the formalism of single
scalar inflation in section 2. We also construct an invariant
extension of the $\zeta$--$\zeta$ correlator in section 4 which
should avoid some of the pitfalls laid out in section 3. Our
conclusions comprise section 5.

\section{Single Scalar Inflation}

The purpose of this section is to review the formalism of single
scalar inflation. This information is well known to experts but may
be unfamiliar to novices, and laying it out will motivate the
subsequent discussion. We begin by giving the dynamical variables
and their Lagrangian in $D$ spacetime dimensions so as to facilitate
dimensional regularization. Then the classical background is
described. The next step is to define the two perturbation fields,
$\zeta(t,\vec{x})$ and $h_{ij}(t,\vec{x})$ whose correlators give
the scalar and tensor power spectra, respectively. After that we
explain how the gauge is fixed and the constraints are solved to
derive the gauge fixed, constrained Lagrangian. The latter process
can only be carried out perturbatively; we present the quadratic
terms in $\zeta$ and $h_{ij}$. We next discuss the close relation
that exists between the two perturbations and the massless,
minimally coupled scalar, and we exploit this relation to derive
approximate tree order results for the power spectra. The section
closes with a discussion of interactions in the gauge fixed and
constrained Lagrangian.

The dynamical variables of single-scalar inflation are the
$D$-dimensional metric ${\rm g}_{\mu\nu}(t,\vec{x})$ and the
inflaton field $\varphi(t,\vec{x})$. The Lagrangian density is,
\begin{equation}
\mathcal{L} = \frac1{16 \pi G} \, {\rm R} \sqrt{-{\rm g}} -\frac12
\partial_{\mu} \varphi \partial_{\nu} \varphi {\rm g}^{\mu\nu}
\sqrt{-{\rm g}} - V(\varphi) \sqrt{-{\rm g}} \; . \label{Ldef}
\end{equation}
We employ the Arnowitt-Deser-Misner (ADM) decomposition of the
spacetime metric ${\rm g}_{\mu\nu}$ into lapse $N(t,\vec{x})$, shift
$N^i(t,\vec{x})$ and spatial metric $g_{ij}(t,\vec{x})$ \cite{ADM},
\begin{equation}
{\rm g}_{00} \equiv -N^2 \!+\! g_{ij} N^i N^j \quad , \quad {\rm
g}_{0i} \equiv -g_{ij} N^j \quad , \quad {\rm g}_{ij} \equiv g_{ij}
\; .
\end{equation}
Our conventions for the various curvatures are,
\begin{equation}
{\rm R}^{\rho}_{~\sigma\mu\nu} = \partial_{\mu}
\Gamma^{\rho}_{~\nu\sigma} + \Gamma^{\rho}_{~\mu\alpha}
\Gamma^{\alpha}_{~\nu\sigma} - (\mu \leftrightarrow \nu) \quad
,\quad {\rm R}_{\mu\nu} = {\rm R}^{\rho}_{~\mu\rho\nu} \quad , \quad
{\rm R} = {\rm g}^{\mu\nu} {\rm R}_{\mu\nu} \; .
\end{equation}

The background geometry is homogeneous, isotropic and spatially
flat,
\begin{equation}
{\rm g}^0_{\mu\nu} dx^{\mu} dx^{\nu} = -dt^2 + a^2(t) d\vec{x}
\!\cdot\! d\vec{x} \; .
\end{equation}
Derivatives of the scale factor $a(t)$ give the Hubble parameter
$H(t)$and the slow roll parameter $\epsilon(t)$,
\begin{equation}
H(t) \equiv \frac{\dot{a}}{a} \qquad , \qquad \epsilon(t) \equiv
-\frac{\dot{H}}{H^2} \; .
\end{equation}
Another important geometrical quantity is the time $t_k$ at which
the physical wave number $k/a(t)$ of some perturbation equals the
Hubble parameter,
\begin{equation}
k = H(t_k) a(t_k) \; .
\end{equation}
The scalar background is $\varphi_0(t)$. Rather than specifying the
scalar potential $V(\varphi)$ and then solving for $a(t)$ and
$\varphi_0(t)$, it is preferable to regard the scale factor as the
primary quantity and then use the background Einstein equations to
eliminate $\dot{\varphi}_0(t)$ and $V(\varphi_0)$,
\begin{equation}
\dot{\varphi}_0^2 = \frac{(D \!-\!2)}{8 \pi G} \, \epsilon H^2
\qquad , \qquad V(\varphi_0) = \frac{(D \!-\!2)}{16 \pi G} \Bigl[
D\!-\!1 \!-\! \epsilon\Bigr] H^2 \; .
\end{equation}

We follow Maldacena \cite{JM} and Weinberg \cite{SW} in defining the
scalar perturbation $\zeta(t,\vec{x})$ from the determinant of the
spatial metric,
\begin{equation}
\zeta(t,\vec{x}) \equiv \frac1{2(D \!-\!1)} \ln\Bigl( {\rm det}
\Bigl[ g_{ij}(t,\vec{x}) \Bigr] \Bigr) - \ln\Bigl[ a(t)\Bigr] \; .
\label{zetadef}
\end{equation}
The remaining unimodular part of the metric
$\widetilde{g}_{ij}(t,\vec{x})$ is expressed as the exponential of a
traceless graviton field $h_{ij}(t,\vec{x})$,
\begin{equation}
\widetilde{g}_{ij}(t,\vec{x}) \equiv \Bigl( e^{h(t,\vec{x})}
\Bigr)_{ij} = \delta_{ij} + h_{ij} + \frac12 h_{ik} h_{kj} + \dots
\label{hdef}
\end{equation}
The full spatial metric is,
\begin{equation}
g_{ij}(t,\vec{x}) \equiv a^2(t) e^{2 \zeta(t,\vec{x})}
\widetilde{g}_{ij}(t,\vec{x}) \; .
\end{equation}
The scalar and tensor power spectra are defined (for $D = 4$
spacetime dimensions) as,
\begin{eqnarray}
\Delta^2_{\mathcal{R}}(k) & \equiv & \frac{k^3}{2 \pi^2} \lim_{t \gg
t_k} \int \!\! d^3x \, e^{-i \vec{k} \cdot \vec{x}} \Bigl\langle
\Omega \Bigl\vert \zeta(t,\vec{x}) \zeta(t,\vec{0}) \Bigr\vert
\Omega \Bigr\rangle \; , \qquad \label{DR} \\
\Delta^2_{h}(k) & \equiv & \frac{k^3}{2 \pi^2} \lim_{t \gg t_k} \int
\!\! d^3x \, e^{-i \vec{k} \cdot \vec{x}} \Bigl\langle \Omega
\Bigl\vert h_{ij}(t,\vec{x}) h_{ij}(t,\vec{0}) \Bigr\vert \Omega
\Bigr\rangle \; . \qquad \label{Dh}
\end{eqnarray}

Even though ADM notation is used, Maldacena \cite{JM} and Weinberg
\cite{SW} do not follow the ADM procedure of fixing the gauge by
specifying $N(t,\vec{x})$ and $N^i(t,\vec{x})$. They instead use the
background value $\varphi_0(t)$ of the inflaton to fix the temporal
gauge condition,
\begin{equation}
G_0(t,\vec{x}) \equiv \varphi(t,\vec{x}) - \varphi_0(t) = 0 \; .
\label{G0}
\end{equation}
And the $(D-1)$ spatial gauge conditions are that the graviton is
transverse \cite{JM,SW},
\begin{equation}
G_i(t,\vec{x}) \equiv \partial_j h_{ij}(t,\vec{x}) = 0 \; .
\label{Gi}
\end{equation}
The constraint equations are then solved to determine the lapse and
shift as nonlocal functionals $N[\zeta,h](t,\vec{x})$ and
$N^i[\zeta,h](t,\vec{x})$. The fully gauge fixed and constrained
Lagrangian is obtained by substituting these solutions into the
original Lagrangian (\ref{Ldef}) and imposing conditions (\ref{G0})
and (\ref{Gi}).

An exact solution exists for the lapse \cite{KOW} but the only known
technique for finding the shift is by recourse to perturbation
theory. Like many perturbative expansions, it quickly becomes
difficult to derive higher order corrections. However, the free
parts are simple enough,
\begin{eqnarray}
\mathcal{L}_{\zeta^2} & = & \frac{(D \!-\!2) \, \epsilon \,
a^{D-1}}{ 16\pi G} \Biggl\{ \dot{\zeta}^2 - \frac1{a^2} \partial_k
\zeta \partial_k \zeta \Biggr\} \; , \label{freezeta} \\
\mathcal{L}_{h^2} & = & \frac{a^{D-1}}{64\pi G} \Biggl\{
\dot{h}_{ij} \dot{h}_{ij} - \frac1{a^2} \partial_k h_{ij} \partial_k
h_{ij} \Biggr\} \; . \label{freeh}
\end{eqnarray}
From (\ref{freezeta}) we see that the free field expansion for
$\zeta(t,\vec{x})$ is $\sqrt{8\pi G/(D-2)}$ times a canonically
normalized scalar whose plane wave mode functions $u_{\zeta}(t,k)$
obey,
\begin{equation}
\ddot{u}_{\zeta} + \Bigl[(D \!-\! 1) H \!+\!
\frac{\dot{\epsilon}}{\epsilon} \Bigr] \dot{u}_{\zeta} +
\frac{k^2}{a^2} u_{\zeta} = 0 \qquad {\rm with} \qquad u_{\zeta}
\dot{u}_{\zeta}^* - \dot{u}_{\zeta} u_{\zeta}^* =
\frac{i}{\epsilon a^{D-1}} \; . \label{zetamodes}
\end{equation}
Expression (\ref{freeh}) implies that each of the $\frac12 (D-3) D$
graviton polarizations is $\sqrt{32 \pi G}$ times a canonically
normalized, massless, minimally coupled scalar. The plane wave mode
function $u(t,k)$ of the massless, minimally coupled scalar obeys,
\begin{equation}
\ddot{u} + (D \!-\! 1) H \dot{u} + \frac{k^2}{a^2} u = 0 \qquad {\rm
with} \qquad u \dot{u}^* - \dot{u} u^* = \frac{i}{a^{D-1}} \; .
\label{hmodes}
\end{equation}
These free field expansions give the tree order results for the
scalar and tensor power spectra (in $D = 4$ spacetime dimensions),
\begin{eqnarray}
\Delta^2_{\mathcal{R}}(k) & = & \frac{k^3}{2 \pi^2} \lim_{t \gg t_k}
\Bigr\{ 4\pi G \times \vert u_{\zeta}(t,k) \vert^2 + O(G^2) \Bigr\}
\; , \label{exactscalar} \\
\Delta^2_{h}(k) & = & \frac{k^3}{2 \pi^2} \lim_{t \gg t_k} \Bigr\{
32\pi G \times 2 \times \vert u(t,k) \vert^2 + O(G^2) \Bigr\} \; .
\label{exacttensor}
\end{eqnarray}

For general $\epsilon(t)$ there is no elementary expression for
either $u(t,k)$ \cite{TW1} or $u_{\zeta}(t,k)$ \cite{TW2}. However,
for constant $\epsilon$ we have,
\begin{eqnarray}
\lefteqn{\dot{\epsilon}(t) = 0 \qquad \Longrightarrow \qquad
u_{\zeta}(t,k) = \frac{u(t,k)}{\sqrt{\epsilon}} } \nonumber \\
& & \hspace{-.2cm} {\rm and} \qquad u(t,k) = \sqrt{\frac{\pi}{4 (1
\!-\! \epsilon) H a^{D-1}}} \, H^{(1)}_{\nu}\Bigl(
\frac{k}{(1\!-\!\epsilon) H a}\Bigr) \; , \; \nu \equiv \frac{D
\!-\! 1 \!-\! \epsilon}{2 (1 \!-\! \epsilon)} \; . \qquad
\label{consteps}
\end{eqnarray}
Constant $\epsilon$ also implies $H a^{\epsilon}$ is constant.
Exploiting this fact and taking $D=4$ gives,
\begin{equation}
D-4 = 0 = \dot{\epsilon} \qquad \Longrightarrow \qquad \lim_{t
\rightarrow \infty} u(t,k) = -i C(\epsilon) \times
\frac{H(t_k)}{\sqrt{2 k^3}} \; . \label{keyeqn}
\end{equation}
The prefactor $C(\epsilon)$ is unity for $\epsilon = 0$ and has the
general form,
\begin{equation}
C(\epsilon) \equiv \frac{ \Gamma(\frac2{1 -
\epsilon})}{\Gamma(\frac1{1 - \epsilon})} \Bigl[ \frac{1 \!-\!
\epsilon}{2^{\epsilon}} \Bigr]^{\frac1{1 - \epsilon}} \; .
\label{C(eps)}
\end{equation}
Constant $\epsilon(t)$ allows further simplification of the power spectra,
\begin{eqnarray}
\dot{\epsilon}(t) = 0 & \Longrightarrow & \Delta^2_{\mathcal{R}} =
C^2(\epsilon) \times \frac{G H^2(t_k)}{\pi \epsilon(t_k)} + O(G^2) \;
, \label{scalarconsteps} \\
\dot{\epsilon}(t) = 0 & \Longrightarrow & \Delta^2_{h} = C^2(\epsilon)
\times \frac{16 G H^2(t_k)}{\pi} + O(G^2) \; .
\label{tensorconsteps}
\end{eqnarray}
Note that the factors of $C(\epsilon)$ cancel in the
tensor-to-scalar ratio,
\begin{equation}
\dot{\epsilon}(t) = 0 \qquad \Longrightarrow \qquad r \equiv
\frac{\Delta^2_{h}}{\Delta^2_{\mathcal{R}}} = 16 \epsilon \; .
\end{equation}
The latest data from the South Pole Telescope implies $r < 0.17$ at
$95\%$ confidence \cite{SPT}, which means $\epsilon < 0.011$. Hence we
conclude that $1/\epsilon > 94$ is a large number.

The graviton propagator $[\mbox{}_{ij} \Delta_{k\ell}](x;x')$ is
proportional to the propagator $i\Delta(x;x')$ of a massless,
minimally coupled scalar,
\begin{equation}
i\Bigl[\mbox{}_{ij} \Delta_{k\ell}\Bigr](x;x') = 32 \pi G \times
\Bigl[ \Pi_{i (k} \Pi_{\ell) j} \!-\! \frac1{D \!-\! 2} \, \Pi_{ij}
\Pi_{k\ell} \Bigr] \times i\Delta(x;x') \; . \label{gravprop}
\end{equation}
Here $\Pi_{ij} \equiv \delta_{ij} - \frac{\partial_i
\partial_j}{\nabla^2}$ is the transverse projection operator. For
the special case of constant $\epsilon(t)$, expression
(\ref{freezeta}) implies a similarly close relation between the
$\zeta$ propagator $i\Delta_{\zeta}(x;x')$ and $i\Delta(x;x')$,
\begin{equation}
\dot{\epsilon}(t) = 0 \qquad \Longrightarrow \qquad
i\Delta_{\zeta}(x;x') = \frac{8 \pi G}{(D \!-\! 2) \epsilon} \times
i\Delta(x;x') \; .
\end{equation}
The massless, minimally coupled scalar has a well-known infrared
problem \cite{FPV,Cliff,Seery} which we regulate by working on
$T^{D-1}$ with radius $L$ and then making the integral approximation
for the mode sum \cite{TWJMPW},
\begin{eqnarray}
\lefteqn{i\Delta(x;x') = \int \!\! \frac{d^{D-1}k}{(2\pi)^{D-1}} \,
\theta(k \!-\! L^{-1}) e^{i \vec{k} \cdot (\vec{x} - \vec{x}')} } \nonumber \\
& & \hspace{3cm} \times \Biggl\{ \theta(t \!-\! t') u(t,k) u^*(t',k)
+ \theta(t' \!-\! t) u^*(t,k) u(t',k) \Biggr\} . \qquad
\label{modesum}
\end{eqnarray}

It is tedious and time-consuming to work out higher order terms in
the expansion of $N^i[\zeta,h]$ which are needed to derive the
interactions of the gauge-fixed and constrained Lagrangian. The
$\zeta^3$ interaction was computed by Maldacena \cite{JM}, and
simple results were obtained for the $\zeta^4$ terms by Seery,
Lidsey and Sloth \cite{SLS}. Paying attention only to the factors of
$\epsilon(t)$ and $\zeta$, these two interactions take the form,
\begin{equation}
\mathcal{L}_{\zeta^3} \sim \epsilon^2 \zeta^3 \qquad , \qquad
\mathcal{L}_{\zeta^4} \sim \epsilon^2 \zeta^4 \; . \label{z34}
\end{equation}
Jarhus and Sloth discussed the next two interactions \cite{JS},
\begin{equation}
\mathcal{L}_{\zeta^5} \sim \epsilon^3 \zeta^5 \qquad , \qquad
\mathcal{L}_{\zeta^6} \sim \epsilon^3 \zeta^6 \; . \label{z56}
\end{equation}
Recently results by Xue, Gao and Brandenberger give the lowest
$\zeta$--graviton interactions \cite{RHB},
\begin{equation}
\mathcal{L}_{\zeta h^2} \sim \epsilon \zeta h^2 \qquad , \qquad
\mathcal{L}_{\zeta^2 h} \sim \epsilon \zeta^2 h \qquad , \qquad
\mathcal{L}_{\zeta^2 h^2} \sim \epsilon \zeta^2 h^2 \; . \label{zhs}
\end{equation}

The pattern which emerges is that any interaction with either $2N$
or $2N-1$ powers of $\zeta$ is suppressed by $N$ powers of $\epsilon$.
{\it There is a reason for this}: it prevents non-Gaussian effects
and loop corrections from being enhanced by the factor of $1/\epsilon
> 94$ associated with each extra $\zeta$ propagator. For example,
consider an $\ell$-loop correction to the $\zeta$--$\zeta$ correlator.
If constructed from just the 4-point interaction $\mathcal{L}_{\zeta^4}$,
it would have $\ell$ vertices and $2\ell + 1$ propagators. Assuming
that powers of $H$ balance the dimensions, we find,
\begin{equation}
\Bigl( \frac{G}{\epsilon} \Bigr)^{2\ell + 1} \times \Bigl(
\frac{\epsilon^2}{G} \Bigr)^{\ell} \times H^{2 \ell + 2} = \Bigl(
\frac{G H^2}{\epsilon} \Bigr) \times \Bigl( G H^2\Bigr)^{\ell} \; .
\end{equation}
Had $\mathcal{L}_{\zeta^4}$ been suppressed by only a single power
of $\epsilon$, each loop would have brought an additional factor of
$1/\epsilon$; had $\mathcal{L}_{\zeta^4}$ been unsuppressed, each
loop would have brought an additional factor of $1/\epsilon^2$.

\section{Ten Principles}

The purpose of this section is to help clarify the debate about
infrared loop corrections to the primordial spectra of single scalar
inflation. We expound ten principles, some of which require little
discussion, but are not less important for that. Our list begins with
the distinction between infrared divergences and infrared growth. We
next turn to invariant extensions of the power spectra which avoid
the former but not the latter. Then three important caveats are
presented which should govern (but do not so far) any nonlinear
extension of the variables whose correlator gives the power spectra.
The section closes with an exhortation to search for secular infrared
dependence where it is most likely to occur.

\subsection{IR divergence differs from IR growth}

An insidious confusion has crept into the literature concerning
infrared corrections to the power spectra. This concerns the failure
to distinguish between infrared {\it divergences} --- which derive
from exactly constant graviton fields --- and infrared finite {\it
secular effects} --- which arise from the continual redshift of
ultraviolet gravitons into the infrared. The former are probably
gauge artifacts but the latter should be real.

The coincidence limit of the graviton propagator (\ref{gravprop}) is
a good venue for studying both effects,
\begin{equation}
i \Bigl[\mbox{}_{ij} \Delta_{k\ell}\Bigr](x;x) = \frac{32 \pi G (D
\!-\! 3) D}{ (D \!-\! 2) (D \!+\! 1)} \Bigl[ \delta_{i (k}
\delta_{\ell) j} \!-\! \frac1{D \!-\! 1} \, \delta_{i j} \delta_{k
\ell } \Bigr] i\Delta(x;x) \; . \label{gravcoinc}
\end{equation}
It is apparent that the graviton propagator inherits both its
infrared divergence and its secular growth from the massless,
minimally coupled propagator (\ref{modesum}),
\begin{equation}
i\Delta(x;x) = \frac{2}{(4 \pi)^{\frac{D-1}2} \Gamma(\frac{D-1}2)}
\int_{L^{-1}}^{\infty} \!\! dk \, k^{D-2} \vert u(t,k)\vert^2 \; .
\end{equation}
For constant $\epsilon(t)$ the mode functions are (\ref{consteps})
and we can change variables to $z = k/[(1-\epsilon)H(t)a(t)]$,
\begin{equation}
\dot{\epsilon}(t) \qquad \Longrightarrow \qquad i\Delta(x;x) =
\frac{[(1 \!-\! \epsilon) H(t)]^{D-2}}{2^D \pi^{\frac{D-3}2}
\Gamma(\frac{D-1}2)} \int_{Z(t)}^{\infty} \!\!\!\! dz \, z^{D-2}
\vert H^{(1)}_{\nu}(z) \vert^2 \; ,
\end{equation}
where $Z(t) \equiv [(1 \!-\! \epsilon) L H(t) a(t)]^{-1}$ and we
recall $\nu = \frac12 (D-1-\epsilon)/(1-\epsilon)$. The next step is
to separate the infrared and ultraviolet parts of the integration,
\begin{equation}
\int_{Z(t)}^{\infty} \!\!\!\! dz = \int_{Z(t)}^{1} \!\!\!\! dz +
\int_1^{\infty} \!\!\!\! dz \; .
\end{equation}
For $\epsilon < 0.011$ only the first term in the power series
expansion of the Hankel function is singular at $z = 0$ so the
secular growth derives from it alone,
\begin{equation}
\int_{Z(t)}^1 \!\!\!\! dz \, z^{D-2} \vert H^{(1)}_{\nu}(z) \vert^2
\longrightarrow \frac{2^{2\nu} \Gamma^2(\nu)}{\pi^2} \times \frac{(1
\!-\! \epsilon)}{(D \!-\! 2) \epsilon} \Biggl\{ \Bigl[ (1 \!-\!
\epsilon) L H(t) a(t)\Bigr]^{\frac{(D-2)\epsilon}{1-\epsilon}}
\!\!\!-\! 1 \Biggr\} .
\end{equation}
We can take $L$ to infinity in the other terms. Now multiply by the
factor of $[H(t)]^{D-2}$ and use the fact that $H(t) a^{\epsilon}(t)
= H_1 a_1^{\epsilon}$ is constant to conclude,
\begin{eqnarray}
\lefteqn{ \dot{\epsilon}(t) = 0 \quad \Longrightarrow \quad
i\Delta(x;x) = \frac{[(1 \!-\! \epsilon) H_1]^{D-2}}{ (4
\pi)^{\frac{D}2}} \frac{2^{2 \nu} \Gamma^2(\nu)}{ \sqrt{\pi}
\Gamma(\frac{D-1}2)} \Biggl\{ \frac{(1 \!-\! \epsilon)}{(D \!-\! 2)
\epsilon} } \nonumber \\
& & \hspace{1cm} \times \Biggl[ \Bigl[(1 \!-\! \epsilon) L H_1
a_1\Bigr]^{\frac{(D-2) \epsilon}{1 - \epsilon}} \!\!\!\! - \Bigl[
\frac{a_1}{a(t)} \Bigr]^{(D-2) \epsilon} \Biggr] + {\rm Constant}
\Bigl[ \frac{H(t)}{H_1}\Bigr]^{D-2} \Biggr\} . \qquad
\label{scalcoinc1}
\end{eqnarray}
Taking $\epsilon$ to zero gives the famous infrared logarithm of de
Sitter \cite{VFLS},
\begin{equation}
\epsilon = 0 \quad \Longrightarrow \quad i\Delta(x;x) =
\frac{H_1^{D-2}}{4 \pi^{\frac{D}2}} \frac{2
\Gamma(\frac{D-1}2)}{\sqrt{\pi}} \Biggl\{ \ln\Bigl[ L H_1 a(t)\Bigr]
+ {\rm Constant} \Biggr\} . \label{scalcoinc2}
\end{equation}

It might seem natural to confuse infrared divergences with secular
growth because the two things are so closely related, however, they
are distinct in a number of important ways. The greatest difference
is that infrared divergences derive from field configurations which
are arbitrarily close to being constant in space and time, whereas
the secular growth results from the continual redshift of modes past
horizon crossing. This means that infrared divergences from
gravitons are likely to be pure gauge, whereas the secular growth
they engender is a physical effect. This has prompted the suggestion
\cite{gauge} that an invariant extension of the $\zeta$--$\zeta$
correlator should be infrared finite. We accept this --- subject to
some important caveats to be mentioned shortly --- but we insist
that this in no way precludes the reality of secular growth.

Another important distinction concerns approximations. Infrared
divergences will be correctly reproduced by techniques which treat
the fields as constant, whereas this would not capture the secular
growth. For example, consider the integral of $k^2/a^2(t)$ times a
coincident propagator, which actually occurs in some schemes
\cite{IRstudies}. For $\epsilon = 0$ the exact result is,
\begin{equation}
\int_{t_1}^t \!\! dt' \, \frac{k^2}{a^2(t')} \times \ln\Bigl[L H
a(t')\Bigr] = \frac{k^2}{2 H a_1^2} \Biggl\{ \ln(L H a_1) \!+\!
\frac12 \!-\! \frac{a_1^2 \ln[LH a(t)]}{a^2(t)} \!-\! \frac{a_1^2}{2
a^2(t)} \Biggr\} . \label{exact}
\end{equation}
Treating the coincident propagator as a constant would correctly
reproduce the infrared divergence,
\begin{equation}
\ln(LH) \times \int_{t_1}^{t} \!\! dt' \frac{k^2}{a^2(t')} =
\frac{k^2 \ln(L H)}{2 H a_1^2} \Biggl\{ 1 - \frac{a_1^2}{a^2(t)}
\Biggr\} . \label{goodapprox}
\end{equation}
However, we would make a very serious error by extracting the
infrared logarithm from the integral,
\begin{equation}
\ln[a(t)] \times \int_{t_1}^t \!\! dt' \frac{k^2}{a^2(t')} =
\frac{k^2 \ln[a(t)]}{2 H a_1^2} \Biggl\{ 1 - \frac{a_1^2}{a^2(t)}
\Biggr\} . \label{badapprox}
\end{equation}
We see from (\ref{exact}) that the integral is dominated by its
lower limit, which precludes the secular growth apparent in the
faulty approximation (\ref{badapprox}).

A final distinction between infrared divergences and secular growth
concerns the way the two things depend upon the slow roll parameter
$\epsilon \equiv -\dot{H}/H^2$. One can see from expression
(\ref{scalcoinc1}) that the infrared divergence of the coincident
propagator is worse as $\epsilon$ increases \cite{RHB}, whereas the
secular growth is maximized for $\epsilon = 0$. Indeed, for
$\epsilon > 0$ the coincidence limit approaches a constant, whereas
it grows without bound for $\epsilon = 0$.

\subsection{The leading IR logs might be gauge independent}

Because the perturbation field $\zeta(t,\vec{x})$ is not gauge
invariant, the $\zeta$--$\zeta$ correlator cannot be gauge
independent. This is part of the reason people have proposed that a
gauge invariant extension of the $\zeta$--$\zeta$ correlator should
be infrared finite. However, it is worth noting that there are
different kinds of spacetime dependence, and just because the
constant part of a gauge fixed Green's function is gauge dependent
does not mean that all the other parts are as well. Because the
secular growth apparent in expressions (\ref{scalcoinc1}) and
(\ref{scalcoinc2}) derives from a physical effect --- the continual
redshift of modes past the Hubble radius --- we suspect that the
leading secular growth terms in the $\zeta$--$\zeta$ correlator are
gauge independent. In this regard it is interesting to note that, in
the de Sitter limit of $\epsilon = 0$, the infrared logarithms of
(\ref{gravcoinc}) and (\ref{scalcoinc2}) agree precisely with those
in the ``spin two'' part of the graviton propagator in the
completely different, de Donder gauge \cite{KMW}.

\subsection{Not all gauge dependent quantities are unphysical}

We have seen that infrared divergences from graviton loops ---
although not temporal growth --- are associated with field
configurations $h_{ij}(t,\vec{x})$ which are nearly constant in
space and time. The fact that an {\it exactly} constant graviton
field is pure gauge motivates the suspicion that the infrared
divergence must be pure gauge. The fact that the $\zeta$--$\zeta$
correlator is certainly afflicted by these infrared divergences
\cite{IRstudies} has led to the suggestion that the spatial gauge
condition (\ref{Gi}) defines an unphysical coordinate system, and
that the infrared divergences would cancel if the $\zeta$--$\zeta$
correlator were extended so as to make it invariant under spatial
coordinate transformations \cite{gauge}. The idea is that graviton
contributions to the fluctuation of $\zeta(t,\vec{x})$ only appear
to be large because the gradual accumulation of nearly constant
field configurations has led to a $\widetilde{g}_{ij}(t,\vec{x})$
which is numerically quite far from $\delta_{ij}$, but still nearly
flat, and hence, nearly gauge equivalent to $\delta_{ij}$.

There is much to be said for this point of view, although we will
later describe some problems with its implementation. The point of
this subsection is just to enjoin some caution about the blanket
condemnation of the $\zeta$--$\zeta$ correlator on account of its
being defined in a special gauge. Just because something is gauge
dependent doesn't mean it is unphysical. For example, sums of
products of the gauge dependent Green's functions of flat space
quantum field theory give the measured rates and cross sections of
the Standard Model. That physical content of these rates and cross
sections did not appear out of nowhere when the gauge dependent
Green's functions were formed into rates and cross sections; it was
obviously present even in the original Green's functions, albeit
mingled with some unphysical effects.

Because the $\zeta$--$\zeta$ correlator would necessarily constitute
part of any nonlinear extension of itself, this correlator must {\it
already} contain some gauge independent and physical information. We
have commented on the possibility that this physical information
might include the leading secular growth factors. The need is for a
reliable way of untangling physical effects from gauge artifacts.

\subsection{Not all gauge invariant quantities are physical}

The point of this subsection is that simply extending the
$\zeta$-$\zeta$ correlator so as to make it invariant under spatial
coordinate transformations is not enough. Just because something is
gauge invariant doesn't mean it is physical. For example, the
operator 1 is perfectly invariant, but it tells us nothing about
primordial perturbations.

Indeed, it is amusing to note that the much-impugned
$\zeta$--$\zeta$ correlator is the expectation value of a nonlocal
invariant operator, as is every gauge fixed Green's function
\cite{TW3}. Given any complete gauge condition, such as (\ref{G0}),
(\ref{Gi}) and the residual conditions implicit in the
$i\varepsilon$ convention for the propagators, one can construct the
field-dependent coordinate transformation $x^{\mu} \rightarrow
{x'}^{\mu}(x)$ which enforces that condition on an arbitrary field
configuration. Let $X^{\mu}[g,\varphi](x)$ represent the inverse of
this field-dependent transformation. Then it is straightforward to
verify the invariance of the components of the transformed metric
\cite{TW4},
\begin{equation}
\frac{\partial X^{\rho}(x)}{\partial x^{\mu}} \frac{\partial
X^{\sigma}(x)}{\partial x^{\nu}} \, g_{\rho\sigma}\Bigl(X(x)\Bigr) =
\frac{\partial {X'}^{\rho}(x)}{\partial x^{\mu}} \frac{\partial
{X'}^{\sigma}(x)}{\partial x^{\nu}} \,
g'_{\rho\sigma}\Bigl(X'(x)\Bigr) \; .
\end{equation}
Further, this quantity is constructed to agree with
the original metric in the fixed gauge,
\begin{equation}
\delta\Bigl[{\rm Gauge\ Condition}\Bigr] \times \frac{\partial
X^{\rho}}{\partial x^{\mu}} \frac{\partial X^{\sigma}}{\partial
x^{\nu}} g_{\rho\sigma}\Bigl(X\Bigr) = \delta\Bigl[{\rm Gauge\
Condition}\Bigr] \times g_{\mu\nu} \; . \label{gaugelocal}
\end{equation}

So the proper criticism of the $\zeta$--$\zeta$ correlator cannot be
that it fails to represent the expectation value of a gauge
invariant operator. It must rather be that the operator whose
expectation value it gives does not describe the measured power
spectrum.

\subsection{Nonlocal ``observables'' can null real effects}

The only sorts of invariant operators in general relativity are
nonlocal. It is very dangerous to allow nonlocal observables because
they can be used to argue that real effects are not present. In fact
it is straightforward to construct a nonlocal functional of the
fields which shows absolutely no effects of interactions. We will
illustrate this in the context of a scalar field $\varphi(x)$ whose
Heisenberg equation of motion is,
\begin{equation}
\mathcal{D} \varphi = I[\varphi] \; .
\end{equation}
Here $\mathcal{D}$ is the linearized kinetic operator and
$I[\varphi]$ is an interaction composed of two or more powers of the
field. For example, a scalar with Lagrangian,
\begin{equation}
\mathcal{L} = -\frac12 \partial_{\mu} \varphi \partial_{\nu} \varphi
g^{\mu\nu} \sqrt{-g} -\frac12 m^2 \varphi^2 \sqrt{-g} -
\frac{\lambda}{4!} \varphi^4 \sqrt{-g} \; ,
\end{equation}
has the following kinetic operator and interaction,
\begin{equation}
\mathcal{D} = \frac1{\sqrt{-g}} \partial_{\mu} \Bigl[ \sqrt{-g} \,
g^{\mu\nu} \partial_{\nu} \Bigr] - m^2 \qquad , \qquad I[\varphi] =
\frac{\lambda}{6} \varphi^3 \; .
\end{equation}

The first step of the construction is to act $1/\mathcal{D}$ (with
any desired boundary conditions) on both sides to obtain the
Yang-Feldman equation \cite{YF},
\begin{equation}
\varphi = \varphi_0 + \frac1{\mathcal{D}} I[\varphi] \; .
\label{YFeqn}
\end{equation}
Here $\varphi_0(x)$ is the ``free field'' which obeys $\mathcal{D}
\varphi_0 = 0$. The usual expansion for the full field $\varphi(x)$
in terms of the free field would result from iterating
(\ref{YFeqn}). For our purposes it is better to express the free
field in terms of the full field,
\begin{equation}
\varphi_0[\varphi] \equiv \varphi - \frac1{\mathcal{D}} I[\varphi]
\; . \label{deadly}
\end{equation}
Equation (\ref{deadly}) defines an explicit nonlocal functional of
the full field which shows absolutely no effect of interactions! By
using the construction backwards it is even possible to relate any
two theories --- such as electromagnetism, general relativity or a
complex scalar field theory --- which have the same numbers of
degrees of freedom.

Field redefinitions such as (\ref{deadly}) would be forbidden in
flat space scattering theory because they change the Borcher's
class. If one rejects invariant variables such as (\ref{gaugelocal})
which reduce to the local fields in some gauge, then there is no
alternative to exploring nonlocal observables. But we must equally
well avoid ridiculous cases such as (\ref{deadly})
--- which might not be so easily recognized as absurd,
especially if one harbors a strong prejudice against some feature of
the interaction. What we need is a relatively simple invariant which
gives a plausible theoretical proxy for what is being measured.

\subsection{Renormalization is crucial and unresolved}

Whatever criticisms can be adduced against invariants
(\ref{gaugelocal}) which become local in a fixed gauge, they possess
an enormous advantage with respect to intrinsically nonlocal
invariants: {\it their ultraviolet divergences can be subtracted
using conventional, BPHZ (Bogoliubov, Parasiuk, Hepp and Zimmerman)
counterterms} \cite{BPHZ}. In contrast, an intrinsically nonlocal
and nonlinear invariant, such as every one which has been proposed
\cite{gauge}, would require composite operator renormalization in
order to remove ultraviolet divergences. {\it There is no general
theory for how to do this in quantum gravity.} So insisting on these
sorts of invariant operators in the interest of controlling the
infrared divergence from graviton loops leaves the ultraviolet
divergence these same loops uncompensated.

To better understand the problem we will describe one of the
extensions proposed for the $\zeta$--$\zeta$ correlator
\cite{gauge}. The idea is to continue determining surfaces of
simultaneity with the temporal gauge condition (\ref{G0}), but to
invariantly fix the length between the two perturbation fields in
this surface using the spatial metric without the scale factor,
\begin{equation}
\widehat{g}_{ij}(t,\vec{x}) \equiv e^{2 \zeta(t,\vec{x})}
\widetilde{g}_{ij}(t,\vec{x}) \qquad \Longrightarrow \qquad
\widehat{\Gamma}^{i}_{~jk} \equiv \frac12 \widehat{g}^{i\ell} \Bigl(
\widehat{g}_{\ell j , k} \!+\! \widehat{g}_{k \ell , j} \!-\!
\widehat{g}_{j k , \ell} \Bigr) \; . \label{spacegeom}
\end{equation}
Instead of the correlator between $\zeta(t,\vec{0})$ and
$\zeta(t,\vec{x})$, the arbitrary point $\vec{x}$ is replaced by the
point a distance $\Vert \vec{V} \Vert$, in the spatial geometry
(\ref{spacegeom}), along the geodesic from $\vec{0}$ in the
direction $\vec{V}$, as measured in the spatial frame field at
$(t,\vec{0})$.

At this point we digress to explain that the spatial frame field at
point is the dreibein field $e_{ia}(x)$, which relates to the
spatial geometry (\ref{spacegeom}) as,
\begin{equation}
\widehat{g}_{ij}(x) = e_{ia}(x) e_{jb}(x) \delta^{ab} \quad , \quad
e^{i}_{~a}(x) \equiv \widehat{g}^{ij}(x) e_{ja}(x) \quad , \quad
e_{i}^{~a}(x) \equiv e_{ia}(x) \; .
\end{equation}
If the local rotational freedom is fixed using the symmetric gauge
condition, $e_{ia}(x) = e_{ai}(x)$, the associated Faddeev-Popov
determinant drops out \cite{RPW1}, and the dreibein is just the
positive square root of the spatial metric,
\begin{equation}
e_{ia}(t,\vec{x}) = e^{\zeta(t,\vec{x})} \Bigl( e^{\frac12 h}
\Bigr)_{ia} = e^{\zeta} \Bigl[ \delta_{ia} \!+\! \frac12 h_{ia}
\!+\! \frac18 h_{ij} h_{ja} \!+\! \dots \Bigr] \; .
\end{equation}

The geodesic $X^i[\widehat{g}](\tau,\vec{V})$ we seek is a
functional of the spatial metric (\ref{spacegeom}), and an ordinary
function of the affine parameter $\tau$ and the initial direction
$\vec{V}$, with $\tau$ derivatives denoted by a dot. It obeys the
geodesic equation,
\begin{equation}
\ddot{X}^i + \widehat{\Gamma}^i_{~ jk}(t,\vec{X}) \dot{X}^j
\dot{X}^k = 0 \; ,
\end{equation}
subject to the initial conditions,
\begin{equation}
X^i(0,V) = 0 \qquad , \qquad \dot{X}^i(0,V) = e^{i}_{~a}(t,\vec{0})
V^a \; .
\end{equation}
This type of operator was employed some decades ago to replace the
gauge-fixed metric with a class of nonlocal operators known as
``Mandelstam Covariants'' from which invariant $N$-point functions
could be defined \cite{TW4}. We can adapt that work to give an
expansion for $X^i[\widehat{g}](\tau,\vec{V})$ in terms of the field
$\chi_{ij}$ comprised of both scalar and tensor perturbations,
\begin{equation}
\chi_{ij} \equiv \widehat{g}_{ij} - \delta_{ij} = h_{ij} + 2 \zeta
\delta_{ij} + \frac12 h_{ik} h_{kj} + 2 \zeta h_{ij} + 2 \zeta^2 \delta_{ij}
+ O\Bigl({\rm cubic}\Bigr) \; .
\end{equation}
As in \cite{TW4}, the letters $A^i$, $B^i$ and $C^i$ denote the
first three terms in the expansion,
\begin{equation}
X^i(\tau,\vec{V}) = A^i(\tau,\vec{V}) + B^i(\tau,\vec{V}) +
C^i(\tau,\vec{V}) + O\Bigl(\chi^3\Bigr) \; .
\end{equation}
The results follow from relations (4.9b-d) of that earlier study
\cite{TW4},
\begin{eqnarray}
A^i(\tau,\vec{V}) & \!\!\!=\!\!\! & V^i \tau \; , \label{Aexp} \\
B^i(\tau,\vec{V}) & \!\!\!=\!\!\! & -\frac12 \chi_{ij}(t,\vec{0})
V^j \tau \!-\! \int_0^{\tau} \!\!\! d\tau_1 \!\! \int_0^{\tau_1}
\!\!\!\! d\tau_2 \, b_{ijk}(t,\tau \vec{V}) V^j V^k \; , \label{Bexp} \\
C^i(\tau,\vec{V}) & \!\!\!=\!\!\! & \frac38 \chi_{ij}(t,\vec{0})
\chi_{jk}(t,\vec{0}) V^k \tau \!+\! \int_0^{\tau} \!\!\! d\tau_1
\!\! \int_0^{\tau_1} \!\!\!\! d\tau_2 \, \chi_{ij}(t,\tau_2 \vec{V})
b_{jk\ell}(t,\tau_2 \vec{V}) V^k V^{\ell} \nonumber \\
& & \hspace{-2cm} -\int_0^{\tau} \!\!\! d\tau_1 \!\! \int_0^{\tau_1}
\!\!\!\! d\tau_2 \Bigl[ b_{ijk , \ell}(t,\tau_2 \vec{V}) V^j V^k
B^{\ell}(\tau_2,\vec{V}) \!+\! 2 b_{ijk}(t,\tau_2 \vec{V}) V^j
\dot{B}^k(\tau_2,\vec{V}) \Bigr] \; , \qquad \label{Cexp}
\end{eqnarray}
where $b_{ijk}$ is the first order term in the expansion of
$\widehat{\Gamma}^{i}_{~jk}$,
\begin{equation}
b_{ijk} \equiv \frac12 \Bigl( \chi_{ij , k} \!+\! \chi_{ki , j}
\!-\! \chi_{jk , i} \Bigr) \; .
\end{equation}

Many proposals for absorbing the infrared divergence from graviton
loops employ $X^i[\widehat{g}](\tau,\vec{V})$ to define an extension
of the scalar power spectrum which does not depend upon the spatial
gauge condition (\ref{Gi}) to fix the separation between the two
fluctuation fields \cite{gauge},
\begin{equation}
\Delta^2_{\mathcal{R}}(k) \longrightarrow \frac{k^3}{2 \pi^2}
\lim_{t \gg t_k} \int \!\! d^3V \, e^{-i \vec{k} \cdot \vec{V}}
\Bigl\langle \Omega \Bigl\vert
\zeta\Bigl(t,\vec{X}[\widehat{g}](1,\vec{V}) \Bigr) \zeta(t,\vec{0})
\Bigr\vert \Omega \Bigr\rangle \; . \label{invDelta}
\end{equation}
The concept of evaluating one operator at a spacetime point which is
itself an operator might cause concern, but it is perfectly well
defined in perturbation theory because the first term (\ref{Aexp})
in the expansion of $X^i$ is a $\comp$-number,
\begin{eqnarray}
\lefteqn{ \zeta\Bigl(t,\vec{X}(1,\vec{V}) \Bigr) = \zeta(t,\vec{V})
+ \zeta_{,i}(t,\vec{V}) B^i(1,\vec{V}) } \nonumber \\
& & \hspace{1.5cm} + \zeta_{,i}(t,\vec{V}) C^i(1,\vec{V}) + \frac12
\zeta_{,ij}(t,\vec{V}) B^i(1,\vec{V}) B^j(1,\vec{V}) + O(\chi^4) \;
. \qquad \label{opexp}
\end{eqnarray}
The one loop correction to (\ref{invDelta}) derives from combining
$\zeta(t,\vec{0})$ times the various terms in the operator expansion
(\ref{opexp}), with enough interaction vertices to reach order $G^2$
once the free field expectation value is taken. This means:
\begin{itemize}
\item{The term $\zeta(t,\vec{V})$ requires either two cubic interaction
vertices or a single quartic vertex;}
\item{The term $\zeta_{,i}(t,\vec{V}) B^i(1,\vec{V})$ requires a
single cubic interaction vertex; and}
\item{Neither $\zeta_{,i}(t,\vec{V}) C^i(1,\vec{V})$ nor
$\zeta_{ij}(t,\vec{V}) B^i(1,\vec{V}) B^j(1,\vec{V})$ requires any
interaction vertices.}
\end{itemize}
Note that the nonlinear extensions of $\zeta(t,\vec{V})$ in
(\ref{opexp}) effectively provide new interaction vertices.

Quite a lot is known about nonlocal composite operators of the type
in (\ref{invDelta}) from a very explicit one loop computation of the
Mandelstam 2-point function \cite{TW4}. In particular, the
ultraviolet properties of their expectation values are worse than
those of local operators. At one loop order in dimensional
regularization, ordinary Green's functions produce only single
factors of $1/(D-4)$, whereas those of a nonlocal composite operator
of the form (\ref{invDelta}) produce two factors of $1/(D-4)$. To
understand the origin of the other divergence it suffices to
consider a composite operator of the form,
\begin{equation}
\zeta(t,\vec{0}) \times \int_0^1 \!\! d\tau \, \zeta(t,\tau \vec{V})
\times \int_0^1 \!\! d\tau' \, \zeta(t,\tau' \vec{V}) \times
\zeta(t,\vec{V}) \; . \label{divproblem}
\end{equation}
Its free field expectation value produces three terms, one of which
is,
\begin{equation}
\int_0^1 \!\! d\tau \, i\Delta_{\zeta}\Bigl(t,\vec{0};t,\tau
\vec{V}\Bigr) \int_0^1 \!\! d\tau' \,
i\Delta_{\zeta}\Bigl(t,\vec{V};t,\tau' \vec{V}\Bigr) \; .
\label{problem}
\end{equation}
Although the integral of a single propagator over a {\it
four}-dimensional volume converges, its integral over a {\it
one}-dimensional region does not. Expression (\ref{problem})
develops separate ultraviolet divergences from the regions near
$\tau = 0$ and $\tau' = 1$. A more complicated analysis shows one
can also get double poles from cubic composites times a single
interaction \cite{TW4},
\begin{equation}
\zeta(t,\vec{0}) \times \int_0^1 \!\! d\tau \, h(t,\tau \vec{V})
\times \zeta(t,\vec{V}) \times \int \!\! d^Dx \, h(x) \zeta(x)
\zeta(x) \; .
\end{equation}

The problem is not just that the usual one loop single-log
ultraviolet divergence gets promoted to a double-log divergence, it
is also that {\it no one understands how to renormalize nonlocal
composite operators.} The theory of {\it local} composite operators
is well understood in renormalizable theories \cite{SW2,IZ}, and
although it has not been much studied for quantum gravity, one can
extend the general ideas \cite{TW5}. The renormalization of a local
composite operator $\mathcal{O}$ is begun by making a list of the
other local composite operators $\mathcal{O}_i$ of the same
dimensionality --- including the factors of $G$ that cause the
canonical field dimensions of $\mathcal{O}_i$ to increase with each
loop --- with which $\mathcal{O}$ is said to ``mix''.
Renormalization is accomplished by adding to $\mathcal{O}$ a linear
combination of the operators with which it mixes,
\begin{equation}
\mathcal{O} \longrightarrow \mathcal{O} + \delta Z_i \mathcal{O}_i
\; .
\end{equation}
The trouble with extending this scheme to a {\it nonlocal} composite
operator such as (\ref{opexp}) is identifying a finite list of local
(or nonlocal) operators with which it mixes. Because (\ref{opexp})
is not local, it technically involves an infinite number of
derivatives. Or if we are to absorb the divergences with other
nonlocal operators, it is not clear which ones should be used. We do
not assert that there is no solution to this problem, only that it
has not been solved to date. And it is a fact that no one has
devised a technique for renormalizing the Mandelstam 2-point
function, even at one loop order, three decades after its first
computation.

It is often implicitly assumed that the ultraviolet problem  must
decouple from the infrared problem. That is known to be true for
gauge-local operators of the form (\ref{gaugelocal}), for which BPHZ
renormalization suffices, but it obviously cannot be asserted for
nonlocal composite operators in the absence of any procedure for
renormalizing them. This is not quibbling; it is reinforced by solid
facts about local composite operators whose renormalization we do
understand in scalar quantum field theories. For example, the two
loop expectation value of the coincident 3-point vertex of Yukawa
theory on de Sitter background manifests an infrared logarithm {\it
multiplied by an ultraviolet divergence} \cite{MW1}. The proper
renormalization of this composite operator through mixing with a
conformal counterterm $\delta \xi \varphi^2 R \sqrt{-g}$ removes
both the ultraviolet divergence and the infrared logarithm. How do
we know this cannot happen in renormalizing nonlocal composite
operators?

Much of our intuition about infrared divergences derives from
renormalizable scalar potential models in which the leading infrared
logarithms (and infrared divergences) at any order can be proved to
be ultraviolet finite \cite{TW6}. But it is known that there can be
ultraviolet divergences even on leading order infrared logarithms
(and divergences) when scalars are permitted to interact with other
fields \cite{PTW}, in quantum gravity \cite{TW7} and in the
nonlinear sigma model \cite{KK}. Fortunately, these divergences
require only a finite number of counterterms at any order, and we
understand how to proceed. The point of this sub-section is that no
one knows how to resolve the ultraviolet problem for nonlocal
composite operators in quantum gravity, nor do we have any assurance
that such a resolution --- if one even exists --- leaves naive
predictions about the infrared unchanged.

\subsection{Extensions involving $\zeta$ must be $\epsilon$-suppressed}

Another important problem concerns the proposed, partially invariant
extensions of the $\zeta$--$\zeta$ correlator: {\it they disrupt the
careful pattern of $\epsilon$-suppression that is apparent in
interactions (\ref{z34}), (\ref{z56}) and (\ref{zhs}).} Recall from
expression (\ref{freezeta}) that the $\zeta$ propagator goes like
$G/\epsilon$. If one employs the $\zeta^4$ vertex, an $\ell$-loop
correction to the $\zeta$--$\zeta$ correlator has $2\ell + 1$
propagators and $\ell$ vertices, giving a correction of the form,
\begin{equation}
\Bigl( \frac{ G H^2}{\epsilon} \Bigr)^{2\ell + 1} \times \Bigl(
\frac{\epsilon^2}{G H^2}\Bigr)^{\ell} = \Bigl( \frac{G
H^2}{\epsilon} \Bigr) \times \Bigl( G H^2 \Bigr)^{\ell} \; .
\label{oldway}
\end{equation}
This means that loop corrections are not $\epsilon$-enhanced.
However, nonlinear extensions of $\zeta(t,\vec{x})$ such as
(\ref{opexp}) essentially add new vertices which are not
$\epsilon$-suppressed. An $\ell$-loop correction which involves
only these terms has just $\ell + 1$ propagators with no
vertices, to give a correction of the form,
\begin{equation}
\Bigl( \frac{G H^2}{\epsilon} \Bigr)^{\ell + 1} = \Bigl( \frac{ G
H^2}{\epsilon} \Bigr) \times \Bigl( \frac{G H^2}{\epsilon}
\Bigr)^{\ell} \; . \label{newway}
\end{equation}
This means that loop corrections are $\epsilon$-enhanced!

Similar results pertain for non-Gaussianity. If one employs the
$\zeta^3$ vertex, the tree order result for the 3-point correlator
has three propagators and one vertex, giving,
\begin{equation}
\Bigl( \frac{G H^2}{\epsilon} \Bigr)^3 \times \Bigl(
\frac{\epsilon^2}{G H^2} \Bigr) = \Bigl( \frac{G H^2}{\epsilon}
\Bigr) \times G H^2 \; . \label{oldFNL}
\end{equation}
If one employs geodesics to fix the physical relation between the
three points in the $\widehat{g}_{ij}$ geometry then the tree order
contribution simply involves two propagators, giving,
\begin{equation}
\Bigl( \frac{G H^2}{\epsilon} \Bigr)^2 = \Bigl( \frac{G
H^2}{\epsilon} \Bigr) \times \Bigl( \frac{G H^2}{\epsilon} \Bigr) \;
. \label{newFNL}
\end{equation}

The counting would be the same if one additionally replaces the
scalar perturbation field $\zeta$ with the more ``geometrical''
spatial curvature,
\begin{equation}
R = \frac{e^{-2\zeta} \widetilde{g}^{ij}}{a^2} \Bigl[ -2(D \!-\! 2)
\widetilde{D}_i \widetilde{D}_j \zeta \!-\! (D \!-\! 2) (D \!-\! 3)
\partial_i \zeta \partial_j \zeta \!+\! \widetilde{R}_{ij} \Bigr] \; .
\label{spaceR}
\end{equation}
(Note that this has actually been proposed \cite{gauge}!) Recall our
convention that quantities with a tilde are constructed using the
unimodular metric $\widetilde{g}_{ij}$ defined in expression
(\ref{hdef}).

In truth, suppression by factors of $G H^2 \ltwid 10^{-10}$ in
expression (\ref{newway}) is ample to keep loop corrections
unobservable at the present time. (Although not perhaps in the
distant future.) The same is true of non-Gaussianity (\ref{newFNL}).
But it is unsettling that an essentially arbitrary convention about
how we measure distances and angles can so completely alter the
results of dynamics which are evident in (\ref{oldway}) and
(\ref{oldFNL}). That could be avoided by eschewing the spatial
curvature (\ref{spaceR}), and basing nonlinear extensions of the
scalar perturbation $\zeta$ on the geometry of the unimodular metric
$\widetilde{g}_{ij}$, rather than $\widehat{g}_{ij} = e^{2 \zeta}
\widetilde{g}_{ij}$.

\subsection{It is important to acknowledge approximations}

Even at a fixed loop order, exact results are unobtainable in
scalar-driven inflation because we lack the mode functions and
propagators for either the scalar inflaton or the massless,
minimally coupled scalar.\footnote{Of course both mode functions are
known for constant $\epsilon = -\dot{H}/H^2$ --- see
(\ref{consteps}) --- but this cannot suffice because inflation never
ends for constant $\epsilon < 1$, and it never begins for constant
$\epsilon > 1$.} Indeed, it is not even possible to give exact
results for the tree order power spectra! All explicit work on loop
corrections must therefore involve some degree of approximation.
Because quantum gravity computations are tedious and time-consuming,
many authors make additional approximations.

We disparage neither the necessity nor the desirability of
appropriate simplification. However, the loop corrections under
consideration are bound to be very small, which means it is
numerically an excellent approximation to neglect them altogether.
That is fine so long as numerical results are desired, but if one
wishes to make exact statements about the presence or absence of
infrared effects then care must be taken to distinguish between
``small'' and ``zero.'' For this it is essential to {\it identify
approximations and make some attempt to understand their
implications, no matter how valid or obvious they seem.} This rule
might appear tedious and pedantic, but we have witnessed shouting
matches occasioned by its breach.

The careful reader has already encountered examples of
approximations which become problematic for certain purposes. In
section 2 we saw that dropping all $\epsilon$-suppressed
interactions eliminates $\zeta$ from the gauge-fixed and constrained
Lagrangian. This does not imply there are no corrections, just that
they are $\epsilon$-suppressed. In sub-section 3.1 we commented on
the folly of confusing time-dependent secular growth factors with
spacetime constant infrared divergences. It is perfectly valid to
extract the latter from an integral such as (\ref{exact}), but
extending this same procedure to the former gives the completely
false result (\ref{badapprox}) that there is logarithmic growth when
the exact result (\ref{exact}) approaches a constant. And
sub-section 3.6 mentioned the potential problems associated with
suppressing ultraviolet effects. It seems ridiculous to be
quarreling over tiny infrared corrections when everything is
dominated by uncontrolled ultraviolet divergence, the resolution of
which may well affect the infrared.

We close this sub-section by commenting on problems that can arise
from failing to discriminate between long wave length and infinite
wave length. Some cosmologists seem to believe that gravitons
disappear after they experience first horizon crossing, only to
reappear, out of nothing, when (and if) they experience second
crossing. This is nonsense. A graviton with $\vec{k} \neq \vec{0}$
cannot be gauged away, no matter how large its physical wave length
becomes. Far super-horizon modes carry only a small energy, but they
do carry some, and there can still be significant effects from
having {\it many} super-horizon modes. It is simple to show that the
occupation number for a single mode in de Sitter is \cite{TW8},
\begin{equation}
N(t,\vec{k}) = \Big( \frac{ H a(t)}{2 k} \Bigr)^2 \; .
\end{equation}
Hence the total energy density from super-horizon gravitons is
\cite{TW8},
\begin{equation}
\rho_{\rm IR} = 2 \times \int \!\! \frac{d^3k}{[2 \pi a(t)]^3} \,
\theta\Bigl( H a(t) \!-\! k\Bigr) \times N(t,\vec{k}) \times
\frac{k}{a(t)} = \frac{H^4}{8 \pi^2} = \frac{G H^2}{3 \pi} \times
\frac{3 H^2}{8 \pi G} \; .
\end{equation}
This is smaller than the energy density of the cosmological constant
by a factor of $G H^2/3\pi$, which means one should expect nonzero
infrared gravitational corrections that are suppressed by the same
factor.

Note also that gravitons carry spin. Unlike energy, spin does not
redshift, so even very infrared gravitons can still interact with
other particles that have spin. That is irrelevant for the scalar
perturbation at one loop order, but it might be relevant at higher
orders. It also seems to explain the curious fact that one loop
corrections to the field strength of a massless fermion grow like
$\ln[a(t)]$ on de Sitter background \cite{MW2}, whereas massless,
minimally coupled scalars experience no growth \cite{KW}.

\subsection{Sub-horizon modes cannot have large IR logs}

We believe that the infrared divergences from graviton loops are
likely to be gauge artifacts, but that secular growth factors are
physical, at least in some cases. We suspect the situation is
similar to that of soft photon corrections to flat space
scattering in which the infrared cutoff in an exclusive amplitude is
replaced by a physical cutoff (involving the detector's energy
resolution) in the associated, inclusive amplitude \cite{SW3}. Under
this analogy, the exclusive and inclusive amplitudes of flat space
scattering would become the original, gauge-fixed version of some
inflationary observable and its properly constructed, invariant
extension, respectively. A plausible rule would be that the former's
dependence upon the infrared cutoff $L$ is replaced in the latter by
the physical scale of the observable at first horizon crossing.

If our conjecture is correct then the replacement for infrared
corrections to the power spectra at comoving wave number $k$ would
be,
\begin{equation}
L \longrightarrow \frac1{H(t_k) a(t_k)} \; .
\end{equation}
The infrared logarithm of the de Sitter case (\ref{scalcoinc2})
would become,
\begin{equation}
\ln\Bigl[ L H_1 a(t)\Bigr] \longrightarrow \ln\Bigl[
\frac{a(t)}{a(t_k)} \Bigr] \; . \label{enhance}
\end{equation}
The largest this can become for a currently observable mode is about
60. This is an enormous enhancement, which might compensate for
suppression by a single factor of $\epsilon < 0.011$. However, it
cannot overcome the suppression all loop corrections suffer from the
loop counting parameter of $G H^2 \ltwid 10^{-10}$. The conclusion
must therefore be that {\it infrared log corrections to the
observable power spectra are bound to be tiny} \cite{SW}. They could
only be observed with a vast increase in our resolving power,
coupled with a unique theory of inflation which allows for precise
determination of the tree order result. Neither advance is beyond
the realm of possibility \cite{21cm}, but they are not likely
to occur soon.

The same considerations apply to any quantity of comoving scale
$\lambda = 2\pi/k$: the best that can be expected from a loop of
infrared gravitons is a fractional correction of about $G H^2
\ln[a(t)/a(t_k)]$. Hence, {\it inflationary gravitons cannot make
significant corrections to anything which is currently subhorizon.}
The point of this sub-section is that it makes more sense to study
infrared corrections to things whose spatial variation we do not
resolve, such as the vacuum energy and particle masses. Once the
restriction $k/a_0 > H_0$ is abandoned, it is obvious that, during
a sufficiently long period of inflation, the infrared enhancement
factor (\ref{enhance}) can become large enough to overcome
suppression by the loop-counting parameter $G H^2 \ltwid 10^{-10}$.
\footnote{When this happens one must take proper account of the
potential for significant stochastic variations in both time and
space \cite{TTW}.}

Many studies have been made of such effects, both from massless,
minimally coupled scalars (distinct from the inflaton) and from
gravitons. Although our primary concern is with gravitons, some of the
scalar models hold great interest because they are fully renormalizable,
free of the gauge issue and because the stochastic technique of
Starobinsky \cite{AAS} enables us to sum the series of leading
infrared logarithms so as to make predictions about the late time
regime after perturbation theory has broken down \cite{SY}. We shall
therefore mention both scalar and graviton infrared effects on de Sitter
background:
\begin{itemize}
\item{In $\phi^4$ theory both, the vacuum energy and the scalar
mass grow \cite{phi4,JMPW};}
\item{In scalar quantum electrodynamics, the vacuum energy falls and
the photon develops a nonzero mass \cite{SQED};}
\item{In Yukawa theory, the vacuum energy falls and the fermion mass
grows \cite{Yukawa,MW1};}
\item{In the nonlinear sigma model, there are subleading infrared
corrections to the stress tensor \cite{KK};}
\item{In Einstein plus Dirac, the fermion field strength grows \cite{MW2};
and}
\item{In pure quantum gravity, the expansion rate seems to slow \cite{TW8}.}
\end{itemize}
Many of these results are at two and three loop orders, and some
include resummations to all orders. Such computational power is lacking
beyond the de Sitter limit of $\epsilon = 0$, but we should mention a
number of studies which find no secular back-reaction in scalar-driven
inflation at one loop order \cite{no1loop}.

\subsection{Spatially constant quantities are observable}

The point of this sub-section is to reinforce the comment we have
just made about the desirability of seeking infrared loop corrections
to things which are perceived as spatially constant. Many cosmologists
seem to believe that a quantity is unobservable unless it has a finite
spatial extent which is within the current horizon. That is nonsense.
Physicists measure many things that possess no spatial variation.
Among them are the vacuum energy, particle masses, Newton's constant
and the various gauge coupling constants. Far from constants being
unobservable, it is believed that current cosmology is largely driven
by a small cosmological constant!

It should also be emphasized that a physical quantity such as a
graviton does not simply disappear when it experiences first horizon
crossing, only to reappear, out of nowhere, after second crossing.
Kinetic energies redshift, which makes them small, not zero. This
energy does something, and the combined effects of many small energies
can be significant. Gravitons also carry spin which does not redshift
at all.

More generally, we call scalar-driven inflation an ``interacting
quantum field theory'' because all the dynamical variables are
ultimately coupled to one another, although it may require many
perturbative interactions to pass between any two. If we subjected
such a field theory to an asymptotic boundary condition, then causality
can sometimes result in certain global degrees of freedom becoming
exactly constant \cite{RW}. However, no asymptotic condition is
enforced in cosmology, which means that every dynamical variable is
coupled to every other one. When a mode undergoes first horizon
crossing, its couplings to sub-horizon modes become small, not zero.
There can still be significant effects if the small coupling to any
one super-horizon mode is compensated by the large number of
super-horizon modes. This may or may not occur, but it would not
represent a violation of the equivalence principle, or any other
principle. And it has been suggested that the vacuum polarization from
the vast ensemble of super-horizon gravitons generated by a long
phase of inflation can modify the effective gravitational field
equations on large scales in phenomenologically useful ways
\cite{nonlocal}.

\section{A New Invariant Power Spectrum}

A good case has been made that the appearance of infrared
divergences in graviton loop corrections to the $\zeta$--$\zeta$
correlator results from employing the spatial gauge condition
(\ref{Gi}) to fix the geometrical relation between observation
points \cite{gauge}. However, that does not require us to measure
$\zeta$ at geodesically related points. As explained in sub-section
3.6, using geodesics leads to new ultraviolet divergences which no
one currently understands how to renormalize. In this section we
will describe a less singular way of geometrically relating the
observation points. We begin by defining the new observable, it is
then expanded to the order necessary for a one loop computation.
Although we do not make a complete computation, we do argue that the
new construction is likely to avoid the extra ultraviolet divergence
associated with geodesic constructions, and we give an explicit
proof that it cancels the infrared divergence in a single graviton
loop.

\subsection{A New Relation between Points}

To avoid altering the pattern of $\epsilon$-suppression, it is
desirable that our geometric relation should involve only the
unimodular metric $\widetilde{g}_{ij}$. Thus we seek a nonlinear
extension of the scalar power spectrum,
\begin{equation}
\Delta^2_{\mathcal{R}}(k) \longrightarrow \frac{k^3}{2 \pi^2}
\lim_{t \gg t_k} \! \int \!\! d^3V \, e^{-i \vec{k} \cdot \vec{V}}
\Bigl\langle \Omega \Bigl\vert
\zeta\Bigl(t,\vec{X}[\widetilde{g}(t)](\vec{V}) \Bigr)
\zeta(t,\vec{0}) \Bigr\vert \Omega \Bigr\rangle \; , \label{goal}
\end{equation}
where $\vec{X}[\widetilde{g}(t)](\vec{V})$ is geometrically related
(using the metric $\widetilde{g}_{ij}$) to the point $\vec{0}$, in a
way that depends on the $\comp$-number parameter $\vec{V}$.

Ultraviolet divergences derive from operators being brought to
coincidence. For noncoincident 1PI (one-particle-irreducible)
functions this occurs when an interaction vertex is integrated over
$D$-dimensional spacetime. Geodesics produce more severe divergences
because they involve integrating along a 1-dimensional path. The
problem can be ameliorated by increasing the dimensionality of the
surface over which graviton fields are integrated to produce the
functional $\vec{X}[\widetilde{g}(t)](\vec{V})$. One obvious way of
doing this involves the Green's function
$G[\widetilde{g}(t)](\vec{x};\vec{y})$ of some scalar differential
operator on the surfaces of simultaneity defined by the temporal
gauge condition (\ref{G0}).

The simplest candidate would seem to be the covariant scalar
Laplacian which we can write as,
\begin{equation}
\triangle \equiv \partial_i \widetilde{g}^{ij}(t,\vec{x}) \partial_j
= \nabla^2 - h_{ij} \partial_i \partial_j + \frac12 h_{ij}
\partial_i h_{jk} \partial_k + O(h^3) \; . \label{SLap}
\end{equation}
The Green's function is defined by the condition,
\begin{equation}
\triangle \, G[\widetilde{g}(t)](\vec{x};\vec{y}) =
\delta^{D-1}(\vec{x} \!-\! \vec{y}) \; . \label{defG}
\end{equation}
For zero graviton field the result is,
\begin{equation}
G[\delta](\vec{x};\vec{y}) = -\frac{ \Gamma(\frac{D-3}2)}{4
\pi^{\frac{D-1}2}} \frac1{ \Vert \vec{x} \!-\! \vec{y} \Vert^{D-3}}
\; . \label{flatG}
\end{equation}
We can invert (\ref{flatG}) to solve for the square of the
coordinate separation,
\begin{equation}
\Vert \vec{x} \!-\! \vec{y} \Vert^2 = \Biggl[ \frac{- 4
\pi^{\frac{D-1}2} }{ \Gamma(\frac{D-3}2)} \,
G[\delta](\vec{x};\vec{y}) \Biggr]^{\frac{-2}{D-3}} \; .
\label{xysq}
\end{equation}
Of course differentiation with respect to $x^i$ gives $2 (x^i -
y^i)$. One defines $\vec{X}[\widetilde{g}(t)](\vec{V})$ for a
general unimodular metric by setting $\vec{y} = \vec{0}$ and solving
for $\vec{X}$ such that,
\begin{equation}
\frac12 \frac{\partial}{\partial X^i} \Biggl[ \frac{ -4
\pi^{\frac{D-1}2} }{ \Gamma(\frac{D-3}2)} \,
G[\widetilde{g}(t)](\vec{X};\vec{0}) \Biggr]^{\frac{-2}{D-3}} =
\widetilde{e}_{ia}(t,\vec{X}) V^a \; . \label{Xdef}
\end{equation}
Here $\widetilde{e}_{ia}$ is the positive square root of the unimodular
metric,
\begin{equation}
\widetilde{e}_{ia} \equiv \Bigl( e^{\frac12 h} \Bigr)_{ia} = \delta_{ia}
+ \frac12 h_{ia} + \frac18 h_{ij} h_{ja} + O(h^3) \; .
\end{equation}

\subsection{Perturbative Expansion}

One loop results require only the first three terms in the graviton
expansion of $\vec{X}[\widetilde{g}(t)](\vec{V})$,
\begin{equation}
X^i[\widetilde{g}(t)](\vec{V}) = \mathcal{A}^i(t,\vec{V}) +
\mathcal{B}^i(t,\vec{V}) + \mathcal{C}^i(t,\vec{V}) + O(h^3) \; .
\label{Xexp}
\end{equation}
Of course the zeroth order term is just $\mathcal{A}^i = V^i$. To
derive $\mathcal{B}^i$ and $\mathcal{C}^i$ we first expand the
scalar Laplacian,
\begin{equation}
\triangle \equiv \nabla^2 + \delta \triangle \qquad \Longrightarrow
\qquad \delta \triangle = -h_{ij} \partial_i \partial_j + \frac12
h_{ij} \partial_i h_{jk} \partial_k - O(h^3) \; .
\end{equation}
We next express the Green's function as the functional inverse of
$\triangle$ acting on a delta function, and then expand,
\begin{eqnarray}
\lefteqn{G[\widetilde{g}(t)](\vec{x};\vec{y}) =
\frac1{\triangle[\widetilde{g}]} \, \delta^{D-1}( \vec{x} \!-\!
\vec{y}) \; , } \\
& & \hspace{-.3cm} = \Biggl\{ \frac1{\nabla^2} \!-\! \frac1{\nabla^2} \,
\delta \triangle \frac1{\nabla^2} \!+\! \frac1{\nabla^2} \, \delta \triangle
\, \frac1{\nabla^2} \, \delta \triangle \, \frac1{\nabla^2} \!-\!
O(\delta \triangle^3) \Biggr\} \, \delta^{D-1}(\vec{x} \!-\! \vec{y})
\; , \qquad \\
& & \hspace{-.3cm} = G[\delta](\vec{x};\vec{y}) - \int \!\! d^{D-1}u \,
G[\delta](\vec{x};\vec{u}) \, \delta \triangle \, G[\delta](\vec{u};\vec{y})
\qquad \nonumber \\
& & \hspace{.5cm} + \int \!\! d^{D-1}u \! \int \!\! d^{D-1}v \,
G[\delta](\vec{x};\vec{u}) \, \delta \triangle \, G[\delta](\vec{u};\vec{v})
\, \delta \triangle \, G[\delta](\vec{v};\vec{y}) - O(h^3) \; . \qquad
\end{eqnarray}

It will simplify the subsequent analysis if we give names to the first and
second fractional corrections of the quantity inside the square brackets of
expression (\ref{Xdef}),
\begin{eqnarray}
\lefteqn{ \frac{-4 \pi^{\frac{D-1}2} }{\Gamma(\frac{D-3}2)} \,
G[\widetilde{g}(t)](\vec{x};\vec{0}) \equiv \frac1{x^{D-3}}
\Biggl\{1 + \beta(t,\vec{x}) + \gamma(t,\vec{x}) + O(h^3) \Biggr\} \; , }
\label{line1} \\
& & \hspace{-.3cm} \beta(t,\vec{x}) = -\frac{\Gamma(\frac{D-3}2)}{4
\pi^{\frac{D-1}2} } \int \!\! d^{D-1}y \, \frac{h_{ij}(t,\vec{y})}{
\Vert \frac{ \vec{x} - \vec{y}}{x} \Vert^{D-3} } \,
\frac{\partial}{\partial y^i} \frac{\partial}{\partial y^j} \,
\frac1{y^{D-3}} \; , \qquad \label{beta} \\
& & \hspace{-.3cm} \gamma(t,\vec{x}) = \frac{\Gamma(\frac{D-3}2)}{8
\pi^{\frac{D-1}2} } \int \!\! d^{D-1}y \, \frac{h_{ij}(t,\vec{y})}{
\Vert \frac{ \vec{x} - \vec{y}}{x} \Vert^{D-3} } \, \frac{\partial}{
\partial y^i} \Biggl[ h_{jk}(t,\vec{y}) \frac{\partial}{\partial y^k} \,
\frac1{y^{D-3}} \Biggr] \nonumber \\
& & \hspace{1.5cm} + \frac{\Gamma^2(\frac{D-3}2)}{16 \pi^{D-1} }
\int \!\! d^{D-1}y \! \int \!\! d^{D-1}z \, \frac{h_{ij}(t,\vec{y})}{ \Vert
\frac{ \vec{x} - \vec{y}}{x} \Vert^{D-3} } \, \Biggl[\frac{\partial}{
\partial y^i} \frac{\partial}{\partial y^j} \frac1{\Vert \vec{y} \!-\! \vec{z}
\Vert^{D-3}} \Biggr] \nonumber \\
& & \hspace{6cm} \times h_{k\ell}(t,\vec{z})
\frac{\partial}{\partial z^k} \frac{\partial}{\partial z^{\ell}} \,
\frac1{z^{D-3} } + O(h^3) \; . \qquad \label{gamma}
\end{eqnarray}
Substituting (\ref{line1}) into relation (\ref{Xdef}) gives,
\begin{eqnarray}
\lefteqn{ \widetilde{e}_{ia}(t,\vec{X}) V^a = X_i \Biggl\{ 1 -
\frac{2 \beta(t,\vec{X})}{D \!-\! 3} + \frac{ (D \!-\! 1)
\beta^2(t,\vec{X}) }{(D \!-\! 3)^2} - \frac{2 \gamma(t,\vec{X})}{D
\!-\! 3} + O(h^3) \Biggr\} } \nonumber \\
& & \hspace{1cm} + X^2 \Biggl\{ 0 - \frac1{D \!-\! 3} \frac{\partial
\beta}{\partial X^i} + \frac{ (D \!-\! 1) \beta}{ (D \!-\! 3)^2}
\frac{\partial \beta}{\partial X^i} - \frac1{D \!-\! 3}
\frac{\partial \gamma}{\partial X^i} + O(h^3) \Biggr\} \; . \qquad
\end{eqnarray}
A few simple rearrangements leads to a form which can be iterated to
generate the perturbative expansion,
\begin{eqnarray}
\lefteqn{ X_i[\widetilde{g}(t)](\vec{V}) =
\widetilde{e}_{ia}(t,\vec{X}) V^a \Biggl\{1 \!+\! \frac{2
\beta(t,\vec{X})}{D \!-\! 3} \!-\! \frac{(D \!-\! 5)
\beta^2(t,\vec{X})}{(D \!-\! 3)^2} \!+\! \frac{2
\gamma(t,\vec{X})}{D \!-\! 3} \!+\! O(h^3) \Biggr\} } \nonumber \\
& & \hspace{1cm} + \frac{X^2}{D \!-\! 3} \Biggl\{ \frac{\partial
\beta(t,\vec{X}) }{\partial X^i} - \beta(t,\vec{X}) \frac{\partial
\beta(t,\vec{X}) }{\partial X^i} + \frac{\partial \gamma(t,\vec{X})
}{\partial X^i} + O(h^3) \Biggr\} \; . \qquad
\end{eqnarray}
It is now straightforward to obtain results for the first and second
order terms in the expansion (\ref{Xexp}) of
$X^i[\widetilde{g}(t)](\vec{V})$,
\begin{eqnarray}
\lefteqn{ \mathcal{B}^i(t,\vec{V}) = \frac12 h_{ij}(t,\vec{V}) V^j +
\frac{\partial}{\partial V^i} \Biggl[ \frac{V^2 \beta(t,\vec{V})}{D
\!-\! 3} \Biggr] \; , } \label{mathB} \\
\lefteqn{ \mathcal{C}^i(t,\vec{V}) = \frac18 h_{ij}(t,\vec{V})
h_{jk}(t,\vec{V}) V^k + \frac12 h_{ij , k}(t,\vec{V}) V^j
\mathcal{B}^k(t,\vec{V}) } \nonumber \\
& & \hspace{-.3cm} + \frac{h_{ij}(t,\vec{V})}{D \!-\! 3} \, V^j
\beta(t,\vec{V}) + \frac2{D \!-\! 3} \frac{\partial
\beta(t,\vec{V})}{\partial V^i} \, V^j \mathcal{B}^j(t,\vec{V}) +
\frac{2 V^i}{(D \!-\! 3)^2} \, \beta^2(t,\vec{V}) \nonumber \\
& & \hspace{.3cm} + \frac{\partial}{\partial V^i} \Biggl[ \frac{V^2
\frac{\partial \beta(t,\vec{V})}{\partial V^j} }{D \!-\! 3} \Biggr]
\mathcal{B}^j(t,\vec{V}) \!-\! \frac{\partial}{\partial V^i} \Biggl[
\frac{V^2 \beta^2(t,\vec{V}) }{2 (D \!-\! 3)} \Biggr] \!+\!
\frac{\partial}{\partial V^i} \Biggl[ \frac{ V^2 \gamma(t,\vec{V})
}{D \!-\! 3} \Biggr] \; . \qquad \label{mathC}
\end{eqnarray}

\subsection{Ultraviolet Behavior}

No complete, dimensionally regulated one loop computation of the
geodesic-based invariants exists \cite{gauge}. That is why the
ultraviolet problem we described in subsection 3.6 has not been
noted previously. Of course the authors of earlier studies were
interested in solving the infrared problem, so they worked only to
the level of approximation needed to capture the leading infrared
effects. The fearsome effort required to obtain a complete result
for the Mandelstam 2-point function at one loop order \cite{TW4}
makes it easy to sympathize with this attitude.

We shall not here attempt to go any further towards computing the
full one loop result from our construction. However, it is easy to
see that the extra divergence associated with geodesics is likely to
be absent. To show this, consider the sort of expression which
arises from two of the first order geodesic corrections
(\ref{Bexp}),
\begin{equation}
\zeta(t,\vec{0}) \times \int_0^{1} \!\! d\tau \, h_{ij}(t,\tau
\vec{V}) \times \int_0^1 \!\! d\tau' \, h_{k\ell}(t,\tau' \vec{V})
\times \zeta(t,\vec{V}) \; . \label{geodprob}
\end{equation}
The new ultraviolet divergence arises because the graviton
propagator from $\tau \vec{V}$ to $\tau' \vec{V}$ diverges too
strongly at $\tau = \tau'$ to be integrable with respect to $\tau$
and $\tau'$. (This divergence comes in addition to the divergence in
the Fourier transform at $\vec{V} = \vec{0}$, just like the
$1/(D-4)^2$ divergences of the Madelstam 2-point function
\cite{TW4}.) In contrast, the essential part of two first order
length corrections (\ref{mathB}) in our construction is,
\begin{eqnarray}
\lefteqn{\zeta(t,\vec{0}) \times \int \!\! d^{D-1}y
\frac{h_{ij}(t,\vec{y})}{ \Vert \vec{V} \!-\! \vec{y} \Vert^{D-3}}
\frac{\partial}{\partial y^i} \frac{\partial}{\partial y^j}
\frac1{y^{D-3}} } \nonumber \\
& & \hspace{4cm} \times \int \!\! d^{D-1}z
\frac{h_{k\ell}(t,\vec{z})}{ \Vert \vec{V} \!-\! \vec{z}
\Vert^{D-3}} \frac{\partial}{\partial z^k} \frac{\partial}{\partial
z^{\ell}} \frac1{z^{D-3}} \times \zeta(t,\vec{V}) \; . \qquad
\label{noprob}
\end{eqnarray}
The same graviton propagator which gives a new ultraviolet
divergence when integrated over a 1-dimensional surface produces a
finite result when integrated over a 3-volume. We therefore expect
only single factors of $1/(D-4)$ at one loop order.

\subsection{Infrared Behavior}

To understand the leading infrared divergence from graviton loops we
can specialize to the case of $h_{ij}(t,\vec{x})$ being constant in
space and time. Because our invariant extension (\ref{goal}) has the
same form (\ref{invDelta}) as those based on geodesics, and because
it has already been checked that the geodesic constructions
eliminate the one loop infrared divergence \cite{gauge}, our
construction will also absorb the one loop infrared divergence
provided $\vec{X}[\widehat{g}](1,\vec{V})$ agrees with
$\vec{X}[\widetilde{g}](\vec{V})$ for constant $h_{ij}(t,\vec{x})$
(and $\zeta = 0$) in $D = 3 +1$ spacetime dimensions. We first
specialize the geodesic expansions (\ref{Bexp}) and (\ref{Cexp}) to
the case of constant $h_{ij}$, $\zeta = 0$ and $D = 3+1$,
\begin{eqnarray}
B^i(1,\vec{V}) & \longrightarrow & -\frac12 h_{ij} V^j - \frac14
h_{ij} h_{jk} V^k + O(h^3) \; , \qquad \\
C^i(1,\vec{V}) & \longrightarrow & + \frac38 h_{ij} h_{jk} V^k +
O(h^3) \; . \qquad
\end{eqnarray}
Hence the most infrared dominant part of the operator geodesic is,
\begin{equation}
X^i(1,\vec{V}) \longrightarrow V^i -\frac12 h_{ij} V^j + \frac18
h_{ij} h_{jk} V^k + O(h^3) \; . \label{theirX}
\end{equation}

To derive the corresponding expansion of our point
$X^i[\widetilde{g}](\vec{V})$ it is necessary to first obtain
results for the quantities $\beta(t,\vec{x})$ and
$\gamma(t,\vec{x})$ defined in expressions (\ref{beta}-\ref{gamma}).
This is, in turn, facilitated by two simple integrals,
\begin{eqnarray}
\lefteqn{\int \!\! d^3y \, \frac1{\Vert \vec{x} \!-\! \vec{y} \Vert}
\, \frac{\partial}{\partial y^i} \frac{\partial}{\partial y^j}
\frac1{y} = -\frac{2 \pi}{x} \Bigl[ \delta^{ij} \!-\!
\widehat{x}^i \widehat{x}^j \Bigr] \; , \label{int1} } \\
\lefteqn{\int \!\! d^3y \, \frac1{\Vert \vec{x} \!-\! \vec{y} \Vert}
\, \frac{\partial}{\partial y^i} \frac{\partial}{\partial y^j} \int
\!\! d^3z \, \frac1{\Vert \vec{y} \!-\! \vec{z}\Vert} \,
\frac{\partial}{\partial z^k} \frac{\partial}{\partial z^{\ell}}
\frac1{z} } \nonumber \\
& & \hspace{5cm} = \frac{6 \pi^2}{x} \Bigl[ \delta^{( ij} \delta^{k
\ell)} \!-\! 2 \delta^{(i j} \widehat{x}^k \widehat{x}^{\ell)} \!+\!
\widehat{x}^i \widehat{x}^j \widehat{x}^k \widehat{x}^{\ell} \Bigr]
\; , \qquad \label{int2}
\end{eqnarray}
where parenthesized indices are symmetrized and we define the radial
unit vector $\widehat{x}^i \equiv x^i/x$. Specializing expressions
(\ref{beta}-\ref{gamma}) to constant (and traceless) $h_{ij}$ and $D
= 3 + 1$ dimensions, and substituting (\ref{int1}-\ref{int2}) gives,
\begin{eqnarray}
\beta(t,\vec{x}) & \longrightarrow & -\frac12 h_{ij} \widehat{x}^i
\widehat{x}^j \; , \qquad \label{betaexp} \\
\gamma(t,\vec{x}) & \longrightarrow & - \frac14 h_{ij} h_{jk}
\widehat{x}^i \widehat{x}^k \!+\! \frac38 (h_{ij} \widehat{x}^i
\widehat{x}^j )^2 \; . \label{gammaexp} \qquad
\end{eqnarray}
Substituting these expansions into expressions
(\ref{mathB}-\ref{mathC}) gives the first and second order
coordinate corrections, specialized to constant $h_{ij}$ and $D =
3+1$,
\begin{eqnarray}
\mathcal{B}^i(t,\vec{V}) & \longrightarrow & -\frac12 h_{ij} V^j \;
, \\
\mathcal{C}^i(t,\vec{V}) & \longrightarrow & +\frac18 h_{ij} h_{jk}
V^k \; . \qquad
\end{eqnarray}
Adding these results in expression (\ref{Xexp}) produces,
\begin{equation}
X^i[\widetilde{g}](\vec{V}) \longrightarrow V^i - \frac12 h_{ij} V^j
+ \frac18 h_{ij} h_{jk} V^k + O(h^3) \; . \label{ourX}
\end{equation}
Because there is precise agreement between (\ref{theirX}) and
(\ref{ourX}), our nonlinear generalization (\ref{goal}) of the
scalar power spectrum is free of infrared divergences from a single
graviton loop.

\section{Conclusions}

The men of genius who created flat space quantum field theory during
the middle decades of the last century had to define observables with
three basic properties:
\begin{itemize}
\item{Infrared finiteness;}
\item{Renormalizability; and}
\item{A reasonable correspondence to what could then be measured.}
\end{itemize}
The fact that quantum field theoretic effects are now being measured in
cosmology, which cannot be described by the old scattering observables,
has confronted this generation of theorists with the same three problems.
This is an opportunity to write on the book of human history, not a
distraction to be disparaged.

It seems to us that the debate on infrared loop corrections would be
elevated by the general adherence to ten principles:
\begin{enumerate}
\item{IR divergence differs from IR growth;}
\item{The leading IR logs might be gauge independent;}
\item{Not all gauge dependent quantities are unphysical;}
\item{Not all gauge invariant quantities are physical;}
\item{Nonlocal ``observables'' can null real effects;}
\item{Extensions involving $\zeta$ must be $\epsilon$-suppressed;}
\item{Renormalization is crucial and unresolved;}
\item{It is important to acknowledge approximations;}
\item{Sub-horizon modes cannot have large IR logs; and}
\item{Spatially constant quantities are observable.}
\end{enumerate}
Many of these points are known and accepted by experts in fundamental
theory. Their absence from the cosmological literature seems to be
responsible for a number of confusions and unfortunate shouting
matches. We thought it might be a service to the community to state
these principles in one place, along with supporting argumentation.

The problems associated with points 6 and 7 have not been noted
before. Nor has anyone suggested the technique of section 4 for
defining an invariant extension of the $\zeta$--$\zeta$ correlator
that is better behaved than proposals which employ geodesics. We
have expanded the field dependent observation point (\ref{Xexp}) to
the order needed for a one loop computation --- see expressions
(\ref{mathB}) and (\ref{mathC}). We have also shown that our
construction eliminates the infrared divergence from a single
graviton loop, the same as geodesic constructions \cite{gauge}. It
would be interesting to see a complete, dimensionally regulated
computation at one loop order.

\centerline{\bf Acknowledgements}

We have profited from conversations on this subject with T. Banks,
R. H. Brandenberger, A. Hebecker, E. O. Kahya, W. Kinney, K. W. Ng,
V. K. Onemli, T. Prokopec, M. Sloth, T. Tanaka, N. C. Tsamis, Y.
Urakawa and G. Veneziano. This work was partially supported by NWO Veni
Project \# 680-47-406, by NSF grant PHY-0855021, and by the Institute
for Fundamental Theory at the University of Florida.


\begin{thebibliography}{99}

\bibitem{Starobinsky} A. A. Starobinsky, JETP Lett. {\bf 30} (1979)
682.

\bibitem{Mukhanov} V. F. Mukhanov and G. V. Chibisov, JETP Lett.
{\bf 33} (1981) 532.

\bibitem{WMAP} E. Komatsu et al., Astrophys. J. Suppl. {\bf 192}
(2011) 18, arXiv:1001.4538.

\bibitem{KOW} E. O. Kahya, V. K. Onemli and R. P. Woodard, Phys.
Lett. {\bf B694} (2010) 101, arXiv:1006.3999.

\bibitem{21cm} S. R. Furlanetto, S. P. Oh and F. H. Briggs, Phys.
Rept. {\bf 433} (2006) 181, astro-ph/0608032.

\bibitem{first21} T. C. Chang, U. L. Pen, K. Bandura and J. B.
Peterson, Nature {\bf 466} (2010) 463, arXiv:1007.3709.

\bibitem{IRstudies} D. Boyanovsky, H. J. de Vega and N. G. Sanchez,
Nucl. Phys. {\bf B747} (2006) 25, astro-ph/0503669; Phys. Rev. {\bf
D72} (2005) 103006, astro-ph/0507596; M. Sloth, Nucl. Phys. {\bf
B748} (2006) 149, astro-ph/0604488; Nucl. Phys. {\bf B775} (2007)
78, hep-th/0612138; A. Biland\v{z}i\'c and T. Prokopec, Phys. Rev.
{\bf D76} (2007) 103507, arXiv:0704.1905; M. van der Meulen and J.
Smit, JCAP {\bf 0711} (2007) 023, arXiv:0707.0842; D. H. Lyth, JCAP
{\bf 0712} (2007) 016, arXiv:0707.0361; D. Seery, JCAP {\bf 0711}
(2007) 025, arXiv:0707.3377; JCAP {\bf 0802} (2008) 006,
arXiv:0707.3378; JCAP {\bf 0905} (2009) 021, arXiv:0903.2788; N.
Bartolo, S. Matarrese, M. Pietroni, A. Riotto and D. Seery, JCAP
{\bf 0801} (2008) 015, arXiv:0711.4263, Y. Urakawa and K. I Maeda,
Phys. Rev. {\bf D78} (2008) 064004, arXiv:0801.0126, A. Riotto and
M. Sloth, JCAP {\bf 0804} (2008) 030, arXiv:0801.1845, K. Enqvist,
S. Nurmi, D. Podolsky and G. I. Rigopoulos, JCAP {\bf 0804} (2008)
025, arXiv:0802.0395, P. Adshead, R. Easther and E. A. Lim, Phys.
Rev. {\bf D79} (2009) 063504, arXiv:0809.4008, J. Kumar, L. Leblond
and A. Rajaraman, JCAP {\bf 1004} (2010) 024, arXiv:0909.2040; C. P.
Burgess, R. Holman, L. Leblond and S. Shandera, JCAP {\bf 1003}
(2010) 033, arXiv:0912.1608, L. Senatore and M. Zaldarriaga, JHEP
{\bf 1012} (2010) 008, arXiv:0912.2734; arXiv:1203.6354;
arXiv:1203.6884; D. Seery, Class. Quant. Grav. {\bf 27} (2010)
124005, arXiv:1005.1649; S. B. Giddings and M. S. Sloth, JCAP {\bf
1101} (2011) 023, arXiv:1005.1056; JCAP {\bf 1007} (2010) 015,
arXiv:1005.3287; Phys. Rev. {\bf D84} (2011) 063528,
arXiv:1104.0002; arXiv:1109.1000; G. L. Pimental, L. Senatore and M.
Zaldarriaga, arXiv:1203.6651.

\bibitem{gauge} Y. Urakawa and T. Tanaka, Prog. Theor. Phys. {\bf
122} (2009) 779, arXiv:0902.3209, Prog. Theor. Phys. {\bf 122}
(2010) 1207, arXiv:0904.4415; Phys. Rev. {\bf D82} (2010) 121301,
arXiv:1007.0468; Prog. Theor. Phys. {\bf 125} (2011) 1067,
arXiv:1009.2947; JCAP {\bf 1105} (2011) 014, arXiv:1103.1251;
C. T. Byrnes, M. Gerstenlauer, A. Hebecker, S.
Nurmi and G. Tasinato, JCAP {\bf 1008} (2010) 006, arXiv:1005.3307,
M. Gerstenlauer, A. Hebecker and G. Tasinato, JCAP {\bf 1106} (2011)
021, arXiv:1102.0560;
Y.  Urakawa, Prog. Theor. Phys. {\bf 126} (2011) 961, arXiv:1105.1078;
D. Chialva and A. Mazumdar, arXiv:1103.1312.

\bibitem{ADM} R. Arnowitt, S. Deser and C. W. Misner, Phys. Rev. {\bf 116}
(1959) 1322; Phys. Rev. {\bf 117} (1960) 1595; Nuov. Cim. {\bf 15}
(1960) 487; Phys. Rev. {\bf 118} (1960) 1100; J. Math. Phys. {\bf 1}
(1960) 434; Phys. Rev. {\bf 120} (1960) 313; 321; Phys. Rev. {\bf
120} (1960) 321; Ann. Phys. {\bf 11} (1960) 116; Nuov. Cim. {\bf 19}
(1961) 668; Phys. Rev. {\bf 121} (1961) 1556; Phys. Rev. {\bf 122}
(1961) 997; gr-qc/0405109.

\bibitem{JM} J. Maldacena, JHEP {\bf 0305} (2003) 013,
astro-ph/0210603.

\bibitem{SW} S. Weinberg, Phys. Rev. {\bf D72} (2005) 043514,
hep-th/0506236; Phys. Rev. {\bf D74} (2006) 023508, hep-th/0605244.

\bibitem{TW1} N. C. Tsamis and R. P. Woodard, Class. Quant. Grav.
{\bf 20} (2003) 5205, astro-ph/0206010.

\bibitem{TW2} N. C. Tsamis and R. P. Woodard, Class. Quant. Grav.
{\bf 21} (2003) 93, astro-ph/0306602.

\bibitem{SPT} R. Keisler et al., Astrophys. J. {\bf 743} (2011) 88,
arXiv:1105.3182.

\bibitem{FPV} L. H. Ford and L. Parker, Phys. Rev. {\bf D16} (1977) 245;
A. Vilenkin, Nucl. Phys. {\bf B226} (1983) 527.

\bibitem{Cliff} C. P. Burgess, R. Holman, L. Leblond and S. Shandera,
JCAP {\bf 1003} (2010) 033, arXiv:0912.1608.

\bibitem{Seery} D. Seery, Class. Quant. Grav. {\bf 27} (2010) 124005,
arXiv:1005.1649.

\bibitem{TWJMPW} N. C. Tsamis and R. P. Woodard, Class. Quant. Grav. {\bf 11}
(1994) 2969; T. M. Janssen, S. P. Miao, T. Prokopec and R. P.
Woodard, Class. Quant. Grav. {\bf 25} (2008) 245013,
arXiv:0808.2449; S. P. Miao, N. C. Tsamis and R. P. Woodard, J.
Math. Phys. {\bf 51} (2010) 072503, arXiv:1002.4037.

\bibitem{SLS} D. Seery, J. E. Lidsey and M. S. Sloth, JCAP {\bf
0701} (2007) 027, astro-ph/0610210.

\bibitem{JS} P. R. Jarnhus and M. S. Sloth, JCAP {\bf 0802} (2008)
013, arXiv:0709.2708.

\bibitem{RHB} W. Xue, X. Gao and R. Brandenberger, ``Infrared
Divergences in Inflation and Entropy Perturbations,''
arXiv:1201.0768.

\bibitem{VFLS} A. Vilenkin and L. H. Ford, Phys. Rev. {\bf D26}
(1982) 1231; A. D. Linde, Phys. Lett. {\bf B116} (1982) 335; A. A.
Starobinsky, Phys. Lett. {\bf B117} (1982) 175.

\bibitem{KMW} E. O. Kahya, S. P. Miao and R. P. Woodard, J. Math.
Phys. {\bf 53} (2012) 022304, arXiv:1112.4420.

\bibitem{TW3} N. C. Tsamis and R. P. Woodard, Class. Quant. Grav.
{\bf 2} (1985) 841; R. P. Woodard, Class. Quant. Grav. {\bf 10}
(1993) 483.

\bibitem{TW4} N. C. Tsamis and R. P. Woodard, Annals Phys. {\bf 215}
(1992) 96.

\bibitem{YF} C. N. Yang and D. Feldman, Phys. Rev. {\bf 79} (1950)
972.

\bibitem{BPHZ} N. N. Bogoliubov and O. Parasiuk, Acta Math. {\bf 97} (1957)
227; K. Hepp, Commun. Math. Phys. {\bf 2} (1966) 301; W. Zimmermann,
Commun. Math. Phys. {\bf 11} (1968) 1; {\bf 15} (1969) 208.

\bibitem{RPW1} R. P. Woodard, Phys. Lett. {\bf B148} (1984) 440.

\bibitem{SW2} S. Weinberg, {\it The Quantum Theory of Fields}, Vol. II
(Cambridge University Press, 1996), pp. 115-118.

\bibitem{IZ} C. Itzykson and J. B. Zuber, {\it Quantum Field Theory}
(McGraw-Hill, New York, 1980) pp. 399-402, 684.

\bibitem{TW5} N. C. Tsamis and R. P. Woodard, Phys. Rev. {\bf D78} (2008)
028501, arXiv:0708.2004.

\bibitem{MW1} S. P. Miao and R. P. Woodard, Phys. Rev. {\bf D74}
(2006) 044019, gr-qc/0602110.

\bibitem{TW6} N. C. Tsamis and R. P. Woodard, Nucl. Phys. {\bf B724}
(2005) 295, gr-qc/0505115.

\bibitem{PTW} T. Prokopec, N. C. Tsamis and R. P. Woodard, Annals Phys. {\bf
323} (2008) 1324, arXiv:0707.0847.

\bibitem{TW7} N. C. Tsamis and R. P. Woodard, Class. Quant. Grav. {\bf 26}
(2009) 105006, arXiv:0807.5006.

\bibitem{KK} H. Kitamoto and Y. Kitazawa, Phys. Rev. {\bf D83}
(2011) 104043, arXiv:1012.5930; Phys. Rev. {\bf D85} (2012) 044062,
arXiv:1109.4892; arXiv:1203.0391.

\bibitem{TW8} N. C. Tsamis and R. P. Woodard, Annals Phys. {\bf 267}
(1998) 145, hep-ph/9712331.

\bibitem{MW2} S. P. Miao and R. P. Woodard, Class. Quant. Grav. {\bf
23} (2006) 1721, gr-qc/0511140; Phys. Rev. {\bf D74} (2006) 044019,
gr-qc/0602110; Class. Quant. Grav. {\bf 25} (2008) 145009,
arXiv:0803.2377.

\bibitem{KW} E. O. Kahya and R. P. Woodard, Phys. Rev. {\bf D76} (2007)
124005, arXiv:0709.0536; Phys. Rev. {\bf D77} (2008) 084012,
arXiv:0710.5282.

\bibitem{SW3} S. Weinberg, Phys. Rev. {bf 140} (1965) B516.

\bibitem{AAS} A. A. Starobinsky, ``Stochastic de Sitter
(inflationary) stage in the early universe,'' in {\it Field Theory,
Quantum Gravity and Strings}, ed. H. J. de Vega and N. Sanchez
(Springer-Verlag, Berlin, 1986) pp. 107-126.

\bibitem{TTW} N. C. Tsamis, A. Tzetzias and R. P. Woodard, JCAP {\bf 1009}
(2010) 016, arXiv:1006.5681.

\bibitem{SY} A. A. Starobinsky and J. Yokoyama, Phys. Rev.
{\bf D50} (1994) 6357, astro-ph/9407016.

\bibitem{phi4} V. K. Onemli and R. P. Woodard, Class. Quant. Grav. {\bf 19}
(2002) 4607, gr-qc/0204065; Phys. Rev. {\bf D70} (2004) 107301,
gr-qc/0406098; T. Brunier, V. K. Onemli and R. P. Woodard, Class.
Quant. Grav. {\bf 22} (2005) 59, gr-qc/0408080; E. O. Kahya and V.
K. Onemli, Phys. Rev. {\bf D76} (2007) 043512, gr-qc/0612026.

\bibitem{JMPW} T. M. Janssen, S. P. Miao, T. Prokopec and R. P. Woodard,
JCAP {\bf 0905} (2009) 003, arXiv:0904.1151.

\bibitem{SQED} T. Prokopec, O. Tornkvist and R. P. Woodard, Phys. Rev.
Lett. {\bf 89} (2002) 101301, astro-ph/0205331; Annals Phys. {\bf
303} (2003) 251, gr-qc/0205130; T. Prokopec and R. P. Woodard, Am.
J. Phys. {\bf 72} (2004) 60, astro-ph/0303358; Annals Phys. {\bf
312} (2004) 1, gr-qc/0310056.

\bibitem{Yukawa} T. Prokopec and R. P. Woodard, JHEP {\bf 0310} (2003)
059, astro-ph/0309593; B. Garbrecht and T. Prokopec, Phys. Rev. {\bf
D73} (2006) 064036, gr-qc/0602011.

\bibitem{TW9} N. C. Tsamis and R. P. Woodard, Int. J. Mod. Phys.
{\bf D20} (2011) 2847, arXiv:1103.5134, Nucl. Phys. {\bf B474}
(1996) 235, hep-ph/9602315, Ann. Phys. {\bf 253} (1997) 1,
hep-ph/9602316; Phys. Rev. {\bf D54} (1996) 2621, hep-ph/9602317.

\bibitem{no1loop} L. R. Abramo and R. P. Woodard, Phys. Rev. {\bf D65}
(2002) 063515, astro-ph/0109272; G. Geshnizjani and R. H. Brandenberger,
Phys. Rev. {\bf D66} (2002) 123507, gr-qc/0204074; T. Janssen and T.
Prokopec, Annals Phys. {\bf 325} (2010) 948, arXiv:0807.0447;
T. S. Koivisto and T. Prokopec, Phys. Rev. {\bf D83} (2011) 044015,
arXiv:1009.5510.

\bibitem{RW} J. A. Rubio and R. P. Woodard, Class. Quant. Grav.
{\bf 11} (1994) 2225; Class. Quant. Grav. {\bf 11} (1994) 2253,
gr-qc/9312020.

\bibitem{nonlocal} S. Deser and R. P. Woodard, Phys. Rev. Lett.
{\bf 99} (2007) 111301, arXiv:0706.2151;
N. C. Tsamis and R. P. Woodard, Phys. Rev. {\bf D80} (2009) 083512,
arXiv:0904.2368; Phys. Rev. {\bf D81} (2010) 103509, arXiv:1001.4929;
C. Deffayet, G. Esposito-Farese and R. P. Woodard, Phys. Rev.
{\bf D84} (2011) 124054, arXiv:1106.4984.

\end{thebibliography}
\end{document}